\documentclass[a4paper,12pt]{revtex4-1}

\usepackage{graphicx}
\usepackage{tabularx}
\usepackage{dcolumn}
\usepackage{bm}
\usepackage{amsmath,amsopn}
\usepackage{amssymb}
\usepackage{epsfig}
\usepackage{bbm} 

\usepackage{hyperref}
\hypersetup{
    unicode=true,          
    pdftoolbar=true,        
    pdfmenubar=true,        
    pdffitwindow=false,     
    pdfauthor={Carlos Triguero},     
    pdfsubject={Subject},   
    pdfcreator={Creator},   
    pdfproducer={Producer}, 
    pdfkeywords={keywords}, 
    pdfnewwindow=true,      
    colorlinks=true,       
    linkcolor=blue,          
    citecolor=blue,        
    filecolor=blue,      
    urlcolor=blue           
}

\def\Re{\mathbb{R}}

\def\Z{\mathbb{Z}}

\def\Id{{\bf I}}

\def\L{{\cal L}}

\def\bfr{{\bf r}}

\def\m{{\bf m}}

\def\e{{\bf e}}

\def\d{{\bf d}}

\def\s{{\bf s}}

\def\De{\bm{\Delta}}

\def\u{{\bf u}}

\def\v{{\bf v}}

\def\w{{\bf w}}

\def\Q{{\bf Q}}

\def\C{{\bf C}}

\def\K{{\bf K}}

\def\J{{\bf J}}

\def\R{{\bf R}}

\def\F{{\bf F}}  

\def\E{{\bf E}}  

\def\calG{{\cal G}}

\def\calM{{\cal M}}

\begin{document}

\begin{center}
{\bf \Large{Modelling avalanches in martensites}}\\
\vskip20pt
{\bf Francisco J. Perez-Reche}\\
\small{Institute for Complex Systems and Mathematical Biology, SUPA, University of Aberdeen, Aberdeen, AB24 3UE, UK}
\end{center}


\vskip20pt

{\bf Abstract}
Solids subject to continuous changes of temperature or mechanical load often exhibit discontinuous avalanche-like responses. For instance, avalanche dynamics have been observed during plastic deformation, fracture, domain switching in ferroic materials or martensitic transformations. The statistical analysis of avalanches reveals a very complex scenario with a distinctive lack of characteristic scales. Much effort has been devoted in the last decades to understand the origin and ubiquity of scale-free behaviour in solids and many other systems. This chapter reviews some efforts to understand the characteristics of avalanches in martensites through mathematical modelling.

\vskip20pt

\tableofcontents

\makeatletter
\let\toc@pre\relax
\let\toc@post\relax
\makeatother 


%



\section{Introduction}
The physical properties of many materials follow a sequence of abrupt changes when the material is driven by smoothly varying the temperature or an external field. The abrupt changes are referred to as avalanches. Examples of processes exhibiting avalanche dynamics include plastic deformation~\cite{Dimiduk2006,Miguel2001}, fracture~\cite{Petri1994,Baro_PRL2013}, domain switching  in ferroic materials~\cite{Durin_review2004,Salje2014} or martensitic phase transitions~\cite{Vives1994a,PerezRechePRL2005,Chandni_PRL2009,Gallardo_PRB2010,Balandraud_PRB2015}. In fact, avalanche dynamics are not exclusive to driven solids but have been reported for a wide variety of processes including earthquakes~\cite{Ben-Zion_RevGeophy2008_Review}, stock market fluctuations~\cite{Sornette_StockCrash2003}, biological extinctions~\cite{Sole_Nature1997}, epidemics~\cite{Rhodes_Nature1997}, neural dynamics~\cite{Plenz_BookCriticalityNeuralNetworks2014} or motion of animal herds~\cite{Ginelli2015}.

A common feature to all these systems is that avalanches are typically characterised by a remarkable variability with magnitudes that extend over several decades. In most cases, the probability density function $\rho(x)$ for the magnitude $x$ of avalanches is found to be long-tailed with a power-law decay, $\rho(x) \sim x^{-\alpha}$, in a wide interval of $x \in [x_{\text{min}},x_{\text{max}}]$. A pure power-law distribution corresponds to the case $x_{\text{max}}=\infty$ (i.e. a Pareto distribution) and implies a lack of characteristic scales~\cite{Newman_ContempPhys2004_Review}. In this case, the distribution of avalanche magnitudes is the same irrespective of the scale we look at it. Exact scale invariance is never detected in experiments due to several factors which may include limitations of the experimental devices, the finite size of systems or the very nature of the avalanche dynamics which may not be exactly scale-free (for instance, large characteristic events can coexist with power-law distribution~\cite{Gil-Sornette_PRL1996,Cavalcante_PRL2013}). In spite of that, a power-law decay over a wide interval $[x_{\text{min}},x_{\text{max}}]$ implies a large dispersion for $x$ and defining a meaningful scale for the magnitudes of avalanches is essentially impossible (i.e. the statistics may be considered as scale-free in practice). For instance, the mean value $\text{E}[x]$ is typically well defined for experimentally obtained distributions. However, taking $\text{E}[x]$ as a characteristic scale for the magnitude of avalanches is typically meaningless since it is much smaller than the dispersion of the data and is not a representative quantity. 

The large variability of magnitudes observed in non-equilibrium avalanche dynamics is reminiscent of the critical phenomena observed in equilibrium second-order phase transitions~\cite{Stanley1971,Goldenfeld1992,Cardy1996}. Following this similarity, the term criticality is broadly used to refer to scale-free avalanche dynamics~\cite{Bak1997,Sethna2001}. 
Avalanche criticality has indeed been associated with second-order phase transitions for a wide range of systems~\cite{Fisher_PhysRep1998,Sethna2001,Durin_review2004}. However, the mechanisms for robust scale-free avalanche behaviour are still a matter of debate for a number of systems, including martensites~\cite{Chandni_PRL2009,PerezReche2007PRL,Perez-Reche_PRB2016}.

This chapter focuses on modelling of avalanche dynamics in martensitic transformations which are solid to solid first-order phase transitions responsible for unique phenomena such as shape memory and pseudoelasticity~\cite{Otsuka1998SMA} or caloric effects~\cite{Manosa-Planes-Acet_JMaterChemA2013}. In particular, we will mostly deal with avalanches in shape-memory alloys undergoing martensitic transitions from an open crystalline structure to a close-packed structure with lower symmetry~\cite{Christian2002v,Otsuka1998SMA,Ortin-Planes-Delaey_ScienceHysteresis2006}. The martensitic transformation can be induced by decreasing the temperature of the material or by applying a mechanical load. Avalanches in shape-memory alloys have been detected using a range of experimental techniques including  acoustic emission~\cite{Vives1994a,Carrillo1998,PerezRechePRL2005,Baro_PRE2012_PLML}, calorimetry~\cite{Carrillo_PRB1997,Gallardo_PRB2010}, resistivity~\cite{Chandni_PRL2009} or strain imaging~\cite{Balandraud_PRB2015}. Avalanches of similar origin have been also observed in ferroelastic materials with optical microscopy~\cite{Harrison-Salje_APL2010,Harrison-Salje_APL2011} and are also predicted by molecular dynamics simulations~\cite{Salje_PRB2011,Ding-Salje_PRB2013}. 

The specific details of phase transformations may depend on the material composition, the particular sample at hand, etc. However, avalanche statistics and other features are qualitatively similar for different materials. For instance, scale-free avalanches have been reported for many different phase transforming alloys and the value of the power-law exponent $\alpha$ only depends on generic features such as the crystal symmetry of the martensitic phase~\cite{Carrillo1998}. Following this observation, many theoretical studies have relied on toy models that aim for a basic understanding of key universal features exhibited by many systems. 
Within the context of martensites, it is common to use spin models with quenched randomness and metastable dynamics~\cite{Vives1995Universality,Goicoechea1995,Shenoy_PRB2008,Vasseur-Lookman_PRB2010,Sherrington_Chapter2012,Cerruti2008}. Most of these models can be viewed as extensions of prototype spin models such as the zero-temperature Random-Field Ising model (RFIM) which was originally proposed to study avalanches in fluid invasion~\cite{Ji1992} and magnetisation reversal~\cite{Sethna1993}. In Sect.~\ref{Sec:Spin-Models}, we review these models which indeed capture important features of martensites including their hysteresis and scale-free avalanches. 
In contrast, they are not suitable to study mechanically-induced transformations and, for instance, miss the fact that in most shape-memory alloys scale-free avalanches are only observed after a training process which consists in cycling several times through the phase transition~\cite{Carrillo1998,PerezReche2004Cyc,PerezReche2003Cyc}.
During this process, shape-memory alloys develop dislocations~\cite{Krauss_ActaMetall1963,Pons_ActaMetallMater1990,Simon_ActaMater2010,Norfleet_ActaMat2009} which are believed to play an important role in the scale-free character of avalanches in well-trained materials~\cite{PerezReche2007PRL,PerezReche_CMT2009}. In this respect, martensites are essentially different from many other apparently similar systems, say driven ferromagnets displaying Barkhausen noise,
because in those systems training is not required to reach a scale-free behavior. 
Motivated by these facts, spin-like models were developed in Refs.~\cite{PerezReche2007PRL,PerezReche_PRL2008,PerezReche_CMT2009,Perez-Reche_PRB2016}  which are applicable to both mechanically and thermally induced transformations. In addition, they are able to simultaneously handle the phase transition and dislocation activity. In order to unify these two processes, it is necessary to deal with the global symmetry group of crystals which includes large shearing and naturally account for the formation of dislocations~\cite{Ericksen_ArchRationalMechAnal1980,Folkins1991,PitteriZanzotto2003,Conti2004,Bhattacharya2003,Bhattacharya2004}. These arguments are reviewed in Sect.~\ref{Sec:Homogeneous-Deformation} for homogeneous lattices and then extended in Sect.~\ref{Sec:Hetero_Deformation} to deal with heterogeneous deformations which are ubiquitous in martensites. After these general considerations, in Sect.~\ref{Sec:RSSM} we formulate the Random Snap-Spring model (RSSM) which is a minimal framework to study avalanches in both thermally and mechanically driven transformations.  Particular cases of the RSSM are studied in Sections \ref{Sec:ThermalDriven} and \ref{Sec:Mech-driven} for thermally-driven and mechanically-driven transformations, respectively.

\section{Avalanches in spin models}
\label{Sec:Spin-Models}
The zero-temperature Random-Field Ising Model (RFIM) is a prototype model for avalanche dynamics that has been applied to study a wide range of phenomena including magnetisation reversal~\cite{Sethna1993,Sethna_review2004}, fluid invasion~\cite{Ji1992,Koiller2000},  capillary condensation in porous media~\cite{Kierlik2001,Handford_PRE2013b} and even opinion shifts~\cite{Michard-Bouchaud_EPJB2005_RFIM-OpinionDynamics}. The RFIM is defined as a set of $N$ spin variables $\{s_i=\pm1;\;i=1,2,\dots,N\}$ with Hamiltonian~\cite{Sethna_review2004,Urbach1995II}
\begin{equation}
\label{eq:RFIM_Hamiltonian}
{\cal H}=-\sum_{i,j} \left(J_{ij}-\frac{J_{\text{inf}}}{N}\right)s_is_j-H\sum_i s_i-\sum_{i} h_i s_i~.
\end{equation}
Here $\J=\{J_{ij}\}$ is a matrix giving a positive interaction $J_{ij}=J>0$ between nearest neighbours and zero otherwise. The term $J_{\text{inf}}/N$ is an infinite-range interaction which, in the case of magnetic materials, is associated with demagnetising fields~\cite{Urbach1995II}; in martensites, it can be associated with the stiffness of mechanical loading devices (see Sect.~\ref{Sec:Mech-driven}) . The spins are acted globally by an external homogeneous field, $H$, and locally by quenched random fields $h_i$ which are often normally distributed with zero mean and standard deviation $r$. The parameter $r$ gives an effective measure of the degree of disorder.

The system is quasistatically driven by sweeping the field $H$ and thermal fluctuations are neglected. Two different spin flip dynamics have been considered in the past. \emph{Nucleation} dynamics assume that any spin can flip provided this leads to a local decrease of the energy~\cite{Sethna1993}. The other possibility corresponds to \emph{front propagation} dynamics in which spins can only flip if this leads to a decrease of the energy \emph{and} they are neighbouring a previously flipped spin~\cite{Ji1992,Koiller2000}. A necessary condition for a spin to flip in any of the two dynamics is that the flip induces a decrease of the energy. From Eq.~\eqref{eq:RFIM_Hamiltonian}, the energy change associated with the flip of a single spin $s_i$ is $\Delta {\cal H}(s_i \rightarrow -s_i)=2 s_i \hat{f}_i$, where
\begin{equation}
\label{eq:RFIM_LocalField}
\hat{f}_i=\sum_j \left(J_{ij}-\frac{J_{\text{inf}}}{N}\right)s_i+H+h_i~,
\end{equation} 
is the local field acting on $s_i$. A stable configuration $\s=\{s_i\}$ is such that all those spins $s_i$ that are allowed to flip must be aligned with their local field, i.e. $s_i \hat{f}_i>0$. In this case, the energy cannot be decreased by flipping any single spin (the simultaneous flip of more than one spin might lead to a lower energy but this corresponds to other dynamics~\cite{PerezReche2004RFIMField} that will not be considered here). Avalanches start when at least one spin becomes unstable under the driving. The unstable spin flips and may induce other spins to flip, thus generating an avalanche of spin flips which finishes when all the spins in the system are stable again. The number of spins flipped during the avalanche gives the avalanche size, $S$.

Many different numerical and analytical studies have been carried out to characterise the properties of magnetisation ($m=\sum_i s_i/N$) and avalanches in the zero-temperature RFIM. One of the most frequently analysed characteristic of avalanches is their size probability distribution which has been studied as a function of the applied field, $D(S,H)$, and also pooling avalanches observed at any field, $D_{\text{int}}(S)=\int_{-\infty}^{\infty} D(S,H) \text{d}H$. The behaviour predicted by the model depends on the degree of disorder, the dynamics for spin flips and the presence ($J_{\text{inf}}>0$) or absence ($J_{\text{inf}}=0$) of infinite-range interactions. We now summarise the behaviour of the model for each dynamics assuming that $H$ increases monotonically in the interval $[-\infty,\infty]$. 

\subsection{Propagation dynamics}
\label{SubSec:RFIM_Propagation}
In order to study the propagation dynamics, the system is initially prepared with a finite domain of spins flipped up ($s=+1$) and the rest are down ($s=-1$). Upon increasing $H$, the domain of up spins grows and the transformation dynamics reduce to the propagation of the domain boundary in a disordered medium. 
Let us consider the case for $J_{\text{inf}}=0$ first. The domain boundary remains pinned for small values of the applied field and depins at a critical field $H_p(r)$ which is a function of the degree of disorder, $r$. The morphology of the propagating front is self-affine for small disorder, $r<r_{p}$, and self-similar for $r>r_{p}$~\cite{Ji1992,Koiller2000}. The avalanches at the pinning-depinning transition obey a power-law distribution, $D(S,H_p)\sim S^{-\tau_p}$, where the exponent $\tau_p$ depends on the morphology of the propagating front. For $r<r^p$, the critical behaviour belongs to the quenched Edwards-Wilkinson (QEW) universality class of driven self-affine interfaces; the value of the exponent for a 3D system is $\tau_p\simeq 1.3$~\cite{Ji1992,Koiller2000,Durin_review2004}. In the high disorder regime, $r>r^p$,  the avalanche size exponent takes the value $\tau_p\simeq 1.6$ which corresponds to the universality class of isotropic percolation~\cite{Ji1992,Koiller2000}. 

A positive value for infinite-range interaction $J_{\text{inf}}$ brings a restoring force that keeps the domain boundary in the neighbourhood of the pinning-depinning transition, thus providing a feedback mechanism which self-tunes the model to a critical state~\cite{Urbach1995II,Kuntz2000,Carpenter2005}. This phenomenology is interpreted as self-organized criticality. Within the framework of driven elastic interfaces, an analogous self-tuning to a pinning-depinning critical regime is obtained if the interface is driven through a weak spring which provides a feed-back mechanism~\cite{Dickman2000,Alava_2002}. 

\subsection{Nucleation dynamics}
\label{SubSec:RFIM_Nucleation}
\begin{figure}[h]
\includegraphics[width=6cm]{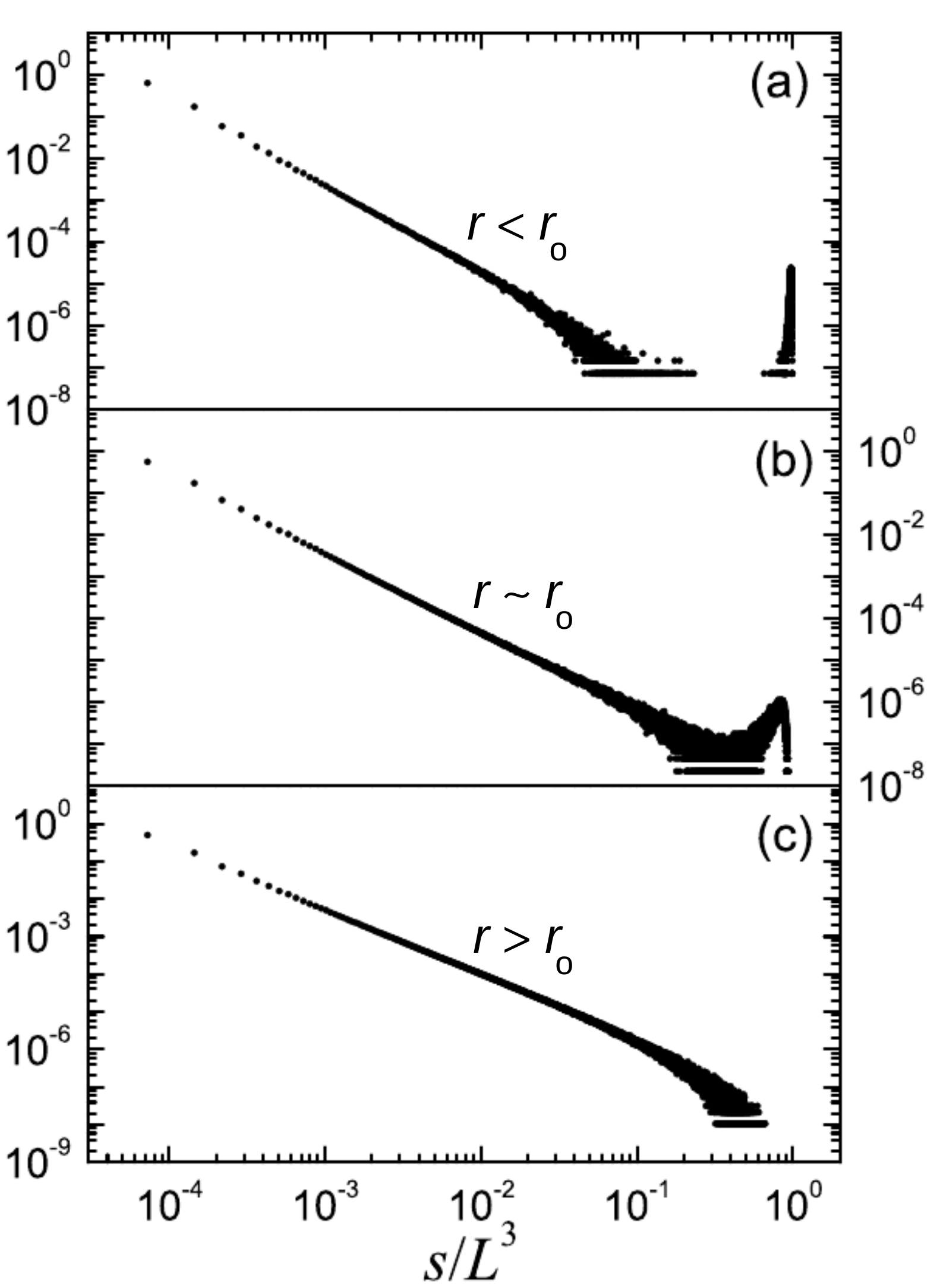}
\caption{Size probability distribution $D_{\text{int}}(S)$ for avalanches observed during the complete transformation in the RFIM with nucleation dynamics. Different panels correspond to different values of disorder:  (a) $r=1.7$, (b) $r=2.21$ and (c) $r=2.6$. The data correspond to a 3D system of linear size $L=24$, i.e. with $N=L^3$ spins  [From \cite{PerezReche2003}, Fig. 1, pg. 134421-4]. }
\label{fig:Avalanche_Size_Distribution_RFIM_PRB2003} 
\end{figure}

Nucleation dynamics allow the flip of any spin that leads to the decrease of the energy. We assume an initial configuration with all the spins in the state $s=-1$ which is stable at the initial field $H=-\infty$. If $J_{\text{inf}}=0$, one observes three different avalanche responses depending on the degree of disorder: pop, crackle and snap~\cite{Sethna2001,PerezReche_PRL2008}. Fig.~\ref{fig:Avalanche_Size_Distribution_RFIM_PRB2003} shows examples of $D_{\text{int}}(S)$ for each regime.
Pop behaviour is observed for large disorder when the transition proceeds through a sequence of many avalanches of small size (see Fig.~\ref{fig:Avalanche_Size_Distribution_RFIM_PRB2003}(c)). In this regime, the magnetisation exhibits a smooth hysteresis loop.  Snap behaviour is observed in systems with disorder smaller than a critical value, $r_o$. In this regime, the magnetisation hysteresis loop exhibits a discontinuity associated with a macroscopically large avalanche that spans the system even in the thermodynamics limit. The peak at large avalanche sizes in Fig.~\ref{fig:Avalanche_Size_Distribution_RFIM_PRB2003}(a) corresponds to the snap event. The pop and snap regimes are separated by a critical disorder ($r_o$) where avalanches have a broad range of sizes. This crackling regime corresponds to an order-disorder (OD) continuous phase transition associated with a $T=0$ critical point $(r_o,H_o)$ in the space spanned by the disorder and driving field~\cite{Sethna1993,Dahmen_PRB1996,Perkovic1999,PerezReche2004RFIMField,Sabhapandit2000,Handford_JSTAT2012,Handford_PRE2013a,Handford_PRE2013c,Balog-Tissier-Tarjus_arxiv2013_Eq-NEq}.  The avalanche size distribution obeys a power law, $D(S,H_o)\sim S^{-\tau_o}$, at the critical point. The exponent $\tau_o$ depends on the dimensionality of the system and, for example, takes the value $\tau_o\simeq 1.6$ in 3D. At the critical disorder, the integrated avalanche size distribution also obeys a power-law decay, $D_{\text{int}}(S) \sim S^{-\tau_o'}$, with an exponent $\tau_o'$ that can be expressed as a function of  $\tau_o$ and other exponents of the system. For a 3D system, $\tau_o' \simeq 2$~\cite{PerezReche2003}. Such a decay is illustrated in Fig.~\ref{fig:Avalanche_Size_Distribution_RFIM_PRB2003}(b); the deviation from a pure power law at large avalanche sizes is a finite size effect which becomes negligible in large systems~\cite{PerezReche2003}.  

For $J_{\text{inf}}>0$, the infinite-range interaction prevents avalanches from growing indefinitely. This implies that an infinite avalanche analogous to the one observed for $r<r_o$ in systems with $J_{\text{inf}}=0$ cannot exist. In contrast, the system self-organises to display front-propagation critical behaviour~\cite{Kuntz2000,Carpenter2005}. The size of avalanches follows a power-law distribution, $D(S)\sim S^{-\tau_p}$, where $\tau_p$ corresponds to the QEW universality class. In general, $\tau_p$ differs from its OD counterpart, $\tau_0$, meaning that infinite-range interactions lead to a change of universality class. In Sect.~\ref{Sec:Mech-driven} we will show that a change in universality class of this type can be induced for mechanically-induced martensitic transformations by varying the stiffness of the loading mechanical device~\cite{PerezReche_PRL2008}.

\subsection{Spin models for martensites}
\label{SubSec:Spins-Martensites}
Several extensions of the zero-temperature RFIM with nucleation dynamics have been proposed to describe martensitic transformations (although avalanches have been only studied in few of them). A key feature of such extensions is the use of spin variables that take more than two values in order to capture the multiple variants of the martensitic phase. The simplest possibility in this direction corresponds to generalisations of the Blume-Emery-Griffiths model with a three-state spin variable, $s \in \{-1,0,1\}$, which allow the transition from austenite ($s=0$) to two variants of martensite ($s=\pm1$) to be described~\cite{Vives1995Universality,Goicoechea1995,Shenoy_PRB2008,Vasseur-Lookman_PRB2010,Sherrington_Chapter2012}. In particular, the avalanche size distribution in thermally-driven martensites was studied in \cite{Vives1995Universality} for a zero-temperature Random-Field Blume-Emery-Griffiths model (RFBEG) with Hamiltonian
\begin{equation}
{\cal H}=-\sum_{i,j} J_{ij}s_is_j-\sum_i h_i s_i^2-g(T) \sum_i s_i^2~,
\end{equation}
where $g(T)$ is a function of temperature that acts as a driving parameter. The fraction of martensite, $q=(1/N) \sum_i s_i^2$, is an increasing function of $g(T)$ so that the austenite and martensite phases are stable for sufficiently positive and negative values of $g(T)$, respectively. In common with the RFIM, the RFBEG  predicts a critical regime with crackling noise associated with an OD critical manifold which separates snap and pop regimes at small and large disorder, respectively.

A Random-Field Potts model with vectorial three-state spins and more realistic dipolarlike interactions than previously considered for the RFBEG was proposed in Ref.~\cite{Cerruti2008}. The model predicts more realistic properties for microstructure and hysteresis than previously obtained with the RFIM or the RFBEG. Criticality is however associated with an OD transition analogous to those exhibited by simpler models.

Phenomenological extensions of the zero-temperature RFIM do not explicitly include elasticity. As a consequence, they are not suitable to describe the behaviour of mechanically-driven martensites. In contrast, mechanically-induced transformations were modelled using a different type of discrete systems consisting of interacting bi-stable elastic units which mechanically behave as \emph{snap-springs}\footnote{A snap-spring in elastic materials can be viewed as the analogue of a spin in magnetic materials.}~\cite{Muller1977,Muller1979,Fedelich1992,Puglisi2000,TV04,Puglisi2006}. Snap-spring models gave interesting insight on the stress-strain hysteresis of martensitic transformations. They were not designed however to capture the emergence of transformation-induced defects and the complexity of avalanches in martensites. Indeed, they traditionally aimed for analytical transparency by assuming simple topologies (e.g. 1D chains) and neglecting spatial heterogeneity. 

\section{Homogeneous deformations of Bravais lattices}
\label{Sec:Homogeneous-Deformation}
The deformation of martensites is highly heterogeneous but, before dealing with such heterogeneities, it is instructive to study the subtleties of homogeneous deformations of crystalline solids~\cite{Ericksen_ArchRationalMechAnal1980,Folkins1991,PitteriZanzotto2003,Conti2004,Bhattacharya2003,Bhattacharya2004}. Let us consider 2D crystals (nets) as a benchmark. Its structure is given by a simple Bravais lattice, defined mathematically as a discrete set of points in the Euclidean space, 
\begin{equation}
\L(\u_a)=\{\R \in \Re^2: \R=n^a \u_a,\; n^a \in \Z \}~.
\end{equation}
Here, the Einstein's summation rule is used. $\R$ gives the position of atoms and $\{n^1,n^2\}$ are the components of $\R$ in the lattice basis defined by the independent pair of vectors $\{\u_1,\u_2\}$. 

\begin{figure}[h]
\centering
\includegraphics[width=11cm]{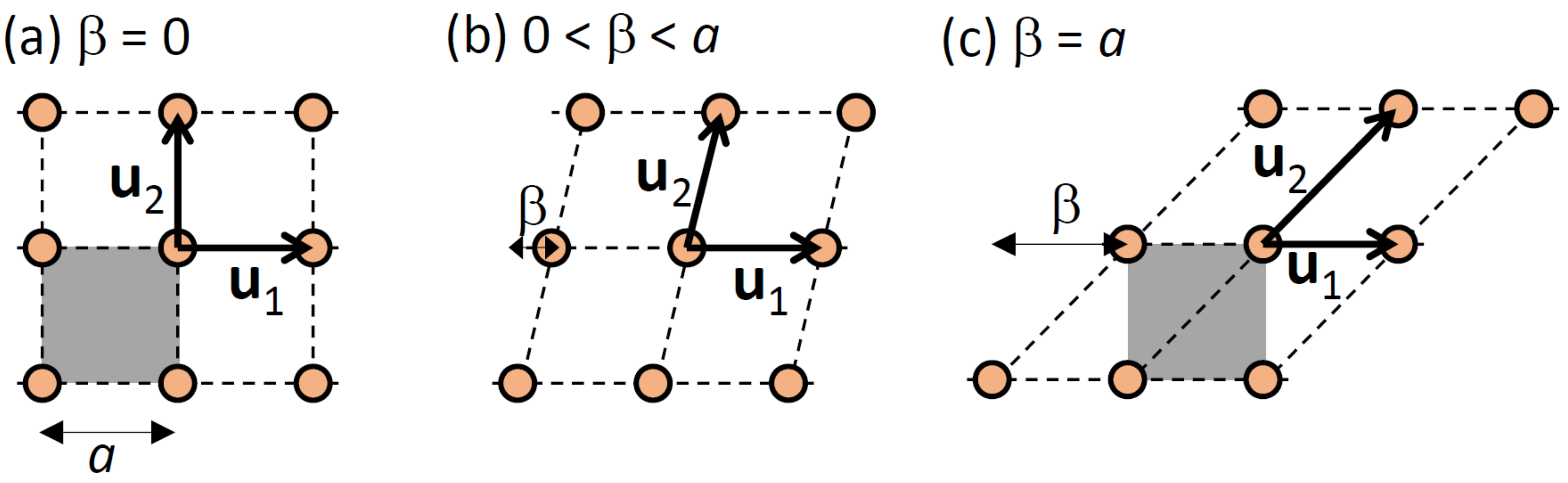}
\caption{Direct shear of a square lattice corresponding to the deformation matrix $\F$ given by Eq.~\eqref{eq:F_square-oblique}. The lattice parameter is $a$. Different panels correspond to different values of the shear parameter $\beta$. The shaded squares in (a) and (c) show that the lattice is identical for $\beta=0$ and $\beta=a$ (assuming that the lattices extend to infinity). In contrast, the lattice basis vectors, $\{\u_1,\u_2\}$, are not identical in (a) and (c).}
\label{fig:Square_Oblique_1} 
\end{figure}

A homogeneous deformation transforms a lattice $\L(\u_a)$ into a new one, $\L(\v_a)$, spanned by basis vectors $\{\v_1,\v_2\}$ which are linearly related to the original lattice vectors, i.e. $\v_a=F_a^b \u_b$. Here,  $\{F_a^b;\; a,b = 1,2\}$ are real-valued elements of an invertible matrix $\F$. Fig.~\ref{fig:Square_Oblique_1} shows a homogeneous deformation of a square lattice by a direct shear of magnitude $\beta$ given by the deformation matrix,
\begin{equation}
\label{eq:F_square-oblique}
\F=\left( \begin{array}{cc}
1 & 0 \\
\beta/a & 1 
\end{array} \right)~,
\end{equation}
where $a$ is the spacing of the square lattice. 

Some deformations lead to new lattices that are identical to the original one, i.e. lattices are symmetric with respect to certain deformations. Such symmetries can be classified into two fundamental types. The first one is the symmetry under orthogonal transformations (i.e. rotations and reflections) and the second is linked to the fact that there are infinitely many different bases describing the same Bravais lattice. We discuss each symmetry type separately in sections \ref{SubSec:Orthogonal-Transformations} and \ref{SubSec:Equivalent-Lattices}. We then use the developed concepts to distinguish between weak and reconstructive transformations and describe the energy of a crystal under homogeneous deformations.

\subsection{Orthogonal transformations}
\label{SubSec:Orthogonal-Transformations}
The relative position of points in Bravais lattices remains unchanged for deformations induced by an orthogonal matrix, i.e. for $\F=\Q \in O(2)$. The physical properties of lattices must then be invariant under such transformations. This means that they must be functions of the lattice metric, $\C$, which is a $2\times2$ Gram matrix with elements given by,
\begin{equation}
\label{eq:C_Definition}
C_{ab}=\u_a^T \u_b~.
\end{equation}
Following this observation, a lattice $\L(\u_a)$ can be defined in terms of the corresponding metric and denoted as $\L(\C)$. All nets can be represented by a point in the metric space $\calM$ spanned by the elements of $\C$. Due to the symmetry $C_{12}=C_{21}$, all 2D nets can in fact be described in terms of three independent matrix elements, $\{C_{11},C_{12},C_{22}\}$.

\begin{figure}[h]
\includegraphics[width=\textwidth]{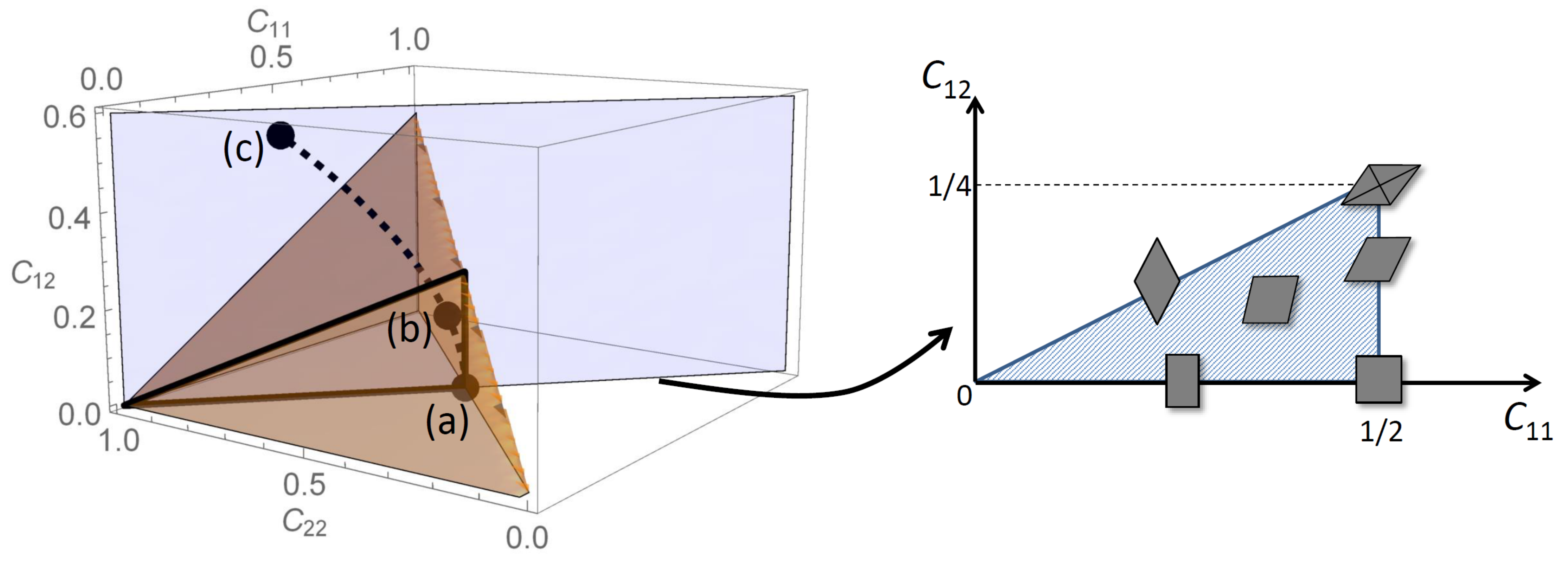}
\caption{Fundamental domain, $FD$, in the lattice metric space $\calM$ spanned by the independent matrix elements, $\{C_{11},C_{12},C_{22}\}$. (Left panel) The brown (dark gray) wedge-shaped region shows the Lagrange $FD$ (Eq.~\eqref{eq.FD}). The dashed line indicates the path associated with the shear deformation given by the matrix in Eq.~\eqref{eq:F_square-oblique}.  The points (a), (b) and (c) correspond to the lattices shown in Fig.~\ref{fig:Square_Oblique_1}. The blue (light gray) plane corresponds to the locus of points satisfying $C_{11}+C_{22}=1$. The thick black triangle shows the boundaries of the region where $FD$ intersects the plane $C_{11}+C_{22}=1$. (Right panel)  The blue (light gray) triangular domain shows the intersection of the $FD$ with the plane $C_{11}+C_{22}=1$ projected on the $(C_{11},C_{12})$ plane. The schematic shapes indicate the different lattice groups: oblique (inside the triangle), rectangular (horizontal line with $C_{12}=0$), fat rhombic (vertical line with $C_{11}=1/2$), skinny rhombic (diagonal line with $C_{12}=\frac{1}{2}C_{11}$) and hexagonal ($C_{11}=1/2$ and $C_{12}=1/4$).}
\label{fig:FD} 
\end{figure}

The six types of Bravais lattices in 2D can be represented in a subset of $\calM$ referred to as a \emph{fundamental domain} ($FD$). A possible choice for $FD$ is the wedge-shaped domain shown in Fig.~\ref{fig:FD}(left) which defines the Lagrange $FD$~\cite{Conti2004}:
\begin{equation}
\label{eq.FD}
FD = \left\{ \C \in \calM, \;\; 0 < C_{11} \leq C_{22}, \;\; 0 \leq C_{12} \leq \frac{C_{11}}{2} \right\}~.
\end{equation}
As shown below, this is just one of the infinitely many possible choices for an $FD$. In Fig.~\ref{fig:FD}(left), the points inside the domain (excluding the boundaries) correspond to oblique (O) lattices.  Rectangular (R) lattices are located on the $C_{12}=0$ plane; square (S) lattices  correspond to the limiting case along the line $C_{11}=C_{22}$. Rhombic (R) lattices are located on the vertical plane along the line with $C_{11}=C_{22}$ and the inclined plane with $C_{11}=2 C_{12}$. Hexagonal (H) lattices are located at the intersection of the latter two planes. The location of different lattice types can be better visualised in the intersection of the $FD$ with the plane $C_{11}+C_{22}=1$ (Fig.~\ref{fig:FD}(right)). 

\subsection{Equivalent lattices}
\label{SubSec:Equivalent-Lattices}
\begin{figure}[h]
\includegraphics[width=5.0cm]{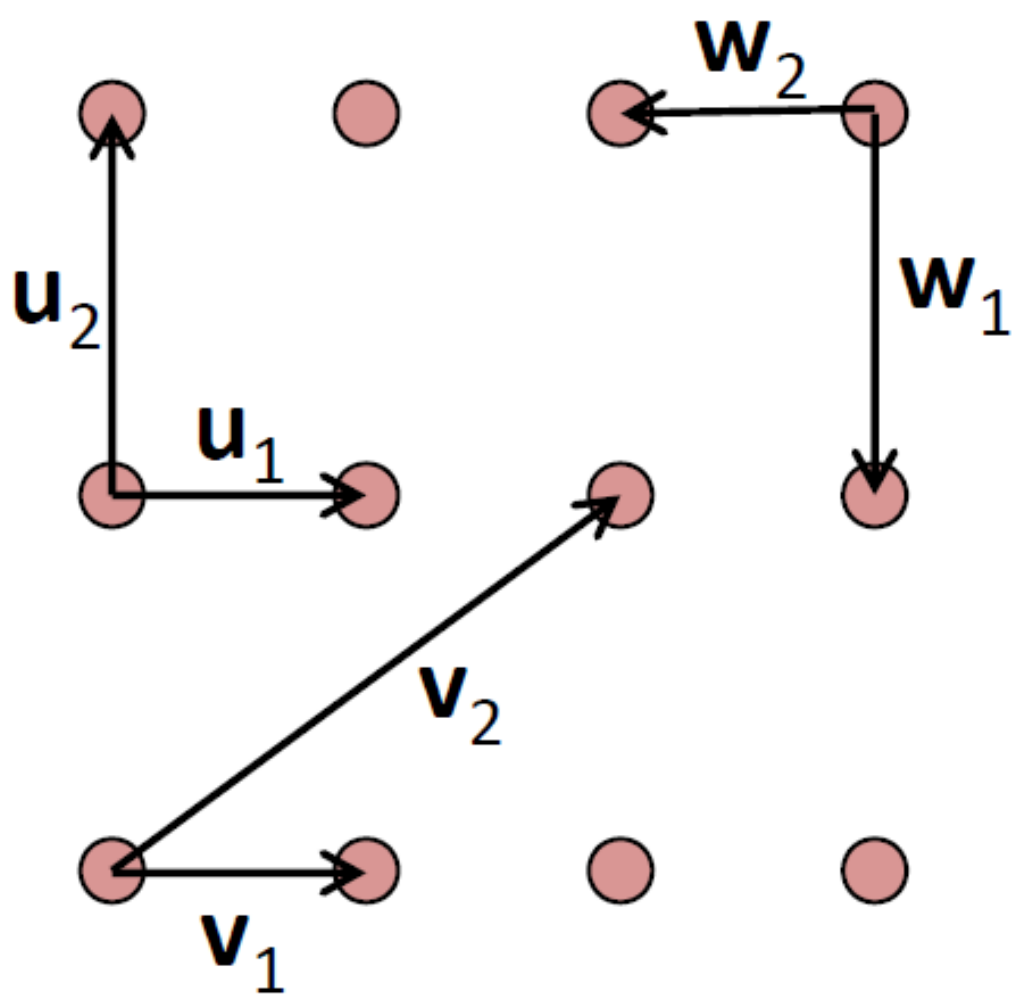}
\caption{Examples of different lattice bases generating a rectangular lattice.}
\label{fig:Base_NonUnique} 
\end{figure}

A lattice base defines a unique Bravais lattice but the contrary is not true. Indeed, there are infinitely many bases which generate the same lattice (Fig.~\ref{fig:Base_NonUnique} shows some examples). The necessary and sufficient condition for two bases, $\u_a$ and $\v_a$, to represent the same lattice is that $\v_a=m_a^{b} \u_b$. Here, $m_a^{b}$ are integer entries of $2 \times 2$ matrices $\m$ with $\operatorname{det} \, \m = \pm 1$, i.e. matrices belonging to the general linear group $GL(2,\Z)$. Fig.~\ref{fig:Base_NonUnique} shows three different bases, $\u_a$, $\v_a$ and $\w_a$  spanning the same rectangular lattice. The bases $\v_a$ and $\w_a$ are related to $\u_a$ by the matrices
\[
\m_v=\left( \begin{array}{cc}
1 & 0 \\
2 & 1 
\end{array} \right) \quad \textrm{and} \quad 
\m_w=\left( \begin{array}{cc}
-1 & 0 \\
0 & -1 
\end{array} \right)~,
\] 
respectively. The metric $\C$ changes as $\C' = \m \C \m^{T}$ under the action of matrices  $\m \in GL(2,\Z)$.

The equivalence of lattices spanned by different basis vectors implies that there are infinitely many copies of each lattice type in $\calM$. For instance, the three different lattice bases shown in Fig.~\ref{fig:Base_NonUnique} give three different copies $\C_u$, $\C_v$ and $\C_w$ in $\calM$ of the same rectangular lattice. 

\begin{figure}[h]
\begin{center}
\includegraphics[width=10cm]{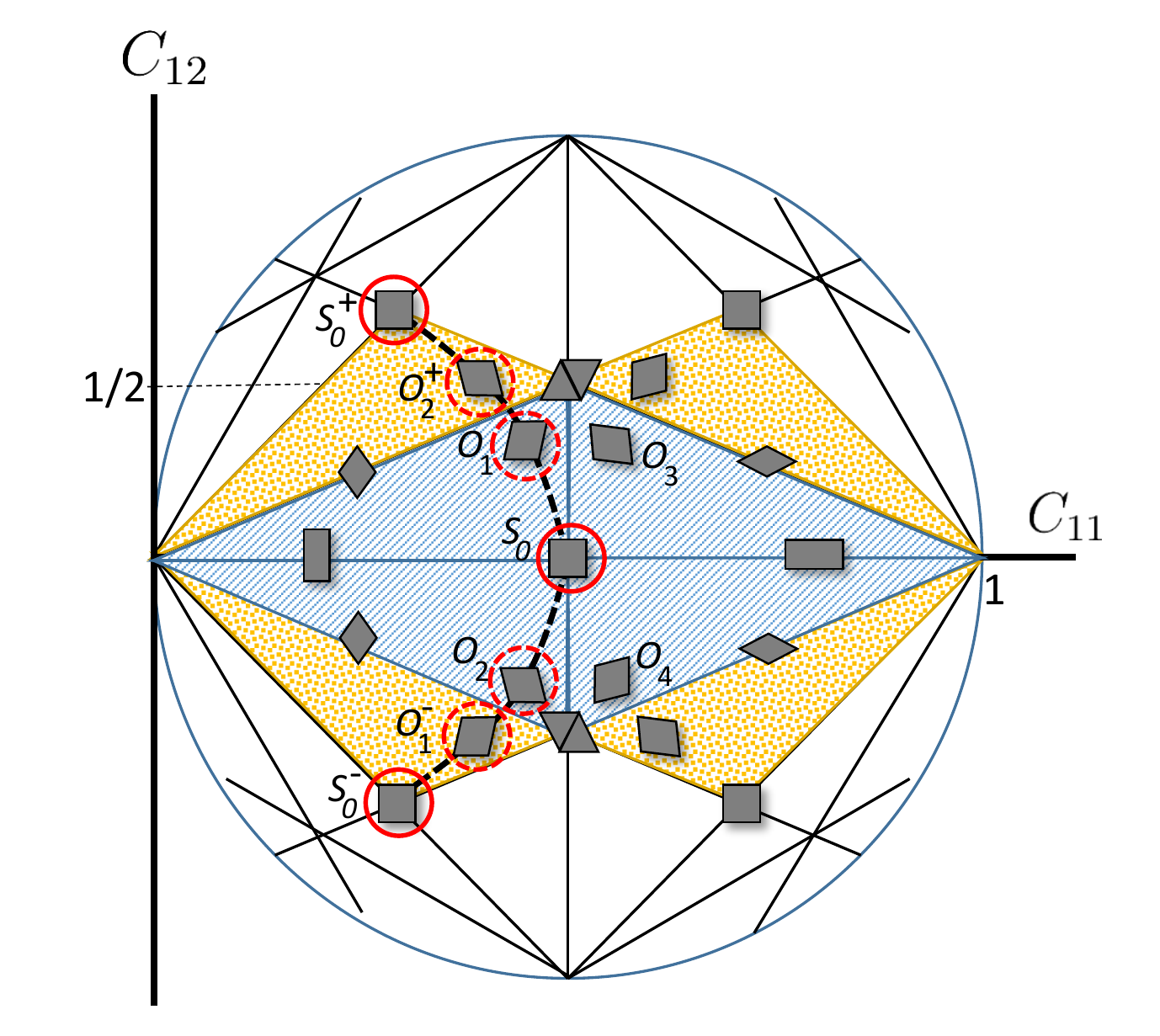}
\end{center}
\caption{Intersection of the metric space $\calM$ with the plane $C_{11}+C_{22}=1$ projected on the $(C_{11},C_{12})$ plane. The dashed domain corresponds to the EPN of the square lattice at $(C_{11},C_{12})=(1/2,0)$. The dotted triangles show copies of the fundamental domain outside the EPN of the square lattice. The dashed thick line shows the path of the lattice under a shear deformation. Energy wells for square austenite and oblique martensite are shown by circles with solid and dashed lines, respectively. The four variants of oblique martensite within the EPN of square $S_0$ are denoted as $O_1-O_4$. $O_1^-$, $O_2^+$ and $S_0^\pm$ are copies of the lattices $O_1$, $O_2$ and $S_0$, respectively, which lay outside the $S_0$ EPN.}
\label{fig:Mspace_Projection_CopiesFD} 
\end{figure}

The concept of fundamental domain introduced in Sect.~\ref{SubSec:Orthogonal-Transformations} can be more rigorously defined as a subset of $\calM$ containing \emph{one and only one} copy of any given lattice under the action of $\m \in GL(2,\Z)$. This means that a given point in a fundamental domain, $FD$, transforms to a different point in the metric space, $\C' = \m \C \m^{T}$, which is necessarily outside $FD$ for any $\m \in GL(2,\Z)$ different from the identity. In the example of Fig.~\ref{fig:Base_NonUnique}, $\C_u$ belongs to the Lagrange $FD$; $\C_v$ and $\C_w$ are outside this domain. Lattice invariance under  $\m \in GL(2,\Z)$ also implies that there are infinitely many copies of a given fundamental domain in the metric space (see Fig.~\ref{fig:Mspace_Projection_CopiesFD}). 

\subsection{Weak and reconstructive transformations}
\label{SubSec:Weak-Reconstructive}
\begin{figure}[h]
\includegraphics[width=6cm]{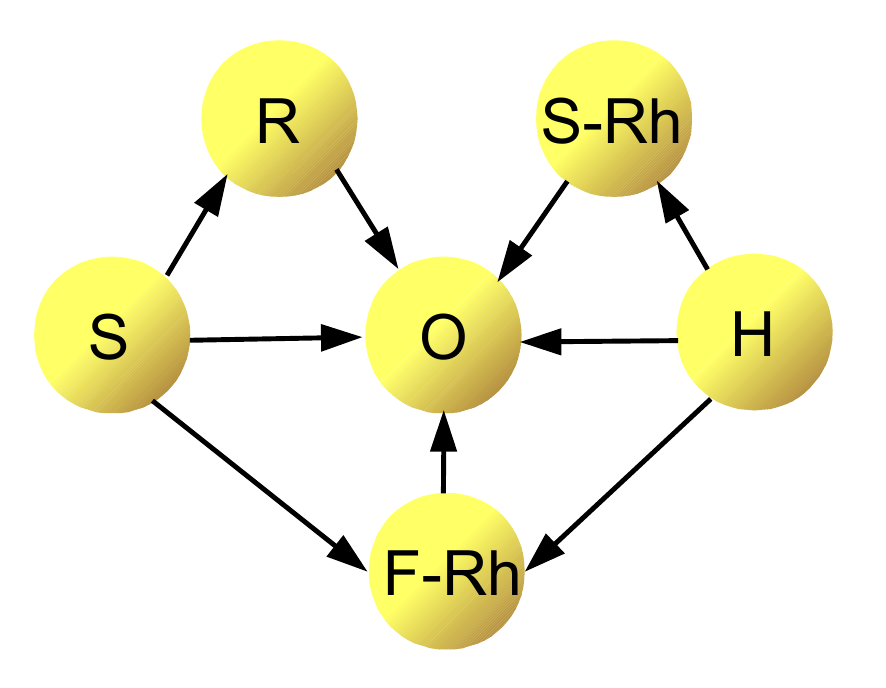}
\caption{Group-subgroup relationships between the six lattice groups of nets (S: square, O: oblique, R: rectangular, F-Rh: fat rhombic, S-Rh: skinny rhombic and H: hexagonal).}
\label{fig:Group-Subgroup} 
\end{figure}

Some matrices $\m \in GL(2,\Z)$ correspond to orthogonal transformations, $\Q \in O(2)$, meaning that they induce a rotation or reflection which transforms the lattice back into itself, i.e. $\C = \m \C \m^{T}$. Such matrices provide a formal definition of the \emph{lattice group} $\calG(\C)$ of any lattice $\L(\C)$. As depicted in Fig.~\ref{fig:FD}(right), there are six different lattice groups for nets~\cite{PitteriZanzotto2003,Conti2004}: square (S), oblique (O), rectangular (R), fat rhombic (F-Rh), skinny rhombic (S-Rh) and hexagonal (H). Lattice groups are finite subgroups of the global symmetry group $GL(2,\Z)$ and, as such, are not sufficient to capture all the symmetries of crystalline solids. In spite of that, a description in terms of only lattice groups can be suitable for \emph{weak} structural phase changes between lattices $\L(\C)$ and $\L(\C^\prime)$ that have a group-subgroup relationship, $\calG(\C) \subseteq \calG(\C^\prime)$. Fig.~\ref{fig:Group-Subgroup} shows the possible structural changes in 2D with a lattice group-subgroup relationship. A number of Landau theories which only account for lattice group symmetries have been proposed to study structural phase transitions in ferroelastic materials and martensites~\cite{Salje1993,Salje_AnnuRevMaterRes2012,Jacobs1985,Shenoy1999,Lookman2003,Ahluwalia2001,Lloveras_PRL2008}. Such theories can successfully describe phase transitions leading to small deformations which are typically the case in ferroelastics.

Indeed, a necessary condition for two lattices $\L(\C)$ and $\L(\C^\prime)$ to have a group-subgroup relationship is that 
$\C^\prime$ corresponds to a sufficiently small deformation of $\C$. More precisely, $\C^\prime$ must be restricted to a sufficiently small open\footnote{Meaning that the boundaries of the neighbourhood do not belong to the neighbourhood itself.} neighbourhood of $\C$ in the space $\calM$. This is the so-called Ericksen-Pitteri neighbourhood (EPN) of $\C$ which was formally introduced in \cite{Ericksen_ArchRationalMechAnal1980,Pitteri1984} (see also \cite{Ball1992,PitteriZanzotto2003,Conti2004,Bhattacharya2004}). 
For example, the dashed region in Fig.~\ref{fig:Mspace_Projection_CopiesFD} is the EPN for the square lattice (the boundaries are excluded). To explore the concept, let us consider again the direct shear of magnitude $\beta$ with deformation matrix given by Eq.~\eqref{eq:F_square-oblique}; the deformed lattice describes the trajectory $\C^\prime(\beta)$ in $\calM$ indicated by the dashed line in Fig.~\ref{fig:FD}(left). For values of $0<\beta<a/2$, the deformed lattice is oblique ($O_1$) and it is a subgroup of the square lattice group (see Fig.~\ref{fig:Group-Subgroup}). In contrast, for $\beta=a/2$, $\C^\prime$ reaches the skinny rhombic lattice group which is not a subgroup of the square lattice. At this point, the system reaches the boundary of the EPN for the original square lattice, $S_0$. Note that the lattice group at the boundary does not have a group-subgroup relationship with the square lattice, thus illustrating the idea that the EPNs are open domains in $\calM$.

Increasing $\beta$ beyond the EPN of $\C$ leads to lattices that cannot be related to lattices in the EPN by elements of the lattice group $\calG(\C)$. Phase transitions inducing large deformations that go beyond the EPN of the parent lattice are \emph{reconstructive}~\cite{ToledanoDmitriev1996} and, strictly speaking, require dealing with the global symmetry group ($GL(2,\Z)$ for nets). Nevertheless, dealing with the global symmetry group is challenging and the global periodicity of the lattice is often approximated using transcendental periodic order parameters~\cite{ToledanoDmitriev1996,Hatch2001,Dmitriev1988}. 

\subsection{The energy of a crystalline solid}
\label{SubSec:Energy-Homogeneous}
The energy of a lattice is a function of the relative position of atoms and the temperature, $T$. For homogeneous lattices, the relative position of atoms is fully determined by the lattice vectors and the energy must be invariant under the symmetry transformations described in sections \ref{SubSec:Orthogonal-Transformations} and \ref{SubSec:Equivalent-Lattices}. For a 2D Bravais lattice, these symmetries imply that the energy is a function $\tilde{\phi}(\C;T)$ of the lattice metric $\C$ (cf. Eq.~\eqref{eq:C_Definition}) which must be invariant under the action of matrices $\m \in GL(2,\Z)$, i.e.
\begin{equation}
\label{eq:GL2Z_phi}
\tilde{\phi}(\C;T)=\tilde{\phi}(\m \C \m^T;T)~.
\end{equation}

At a given temperature, stable lattices correspond to metrics $\C$ that minimise $\tilde{\phi}(\C;T)$. Note that the global symmetry given by Eq.~\eqref{eq:GL2Z_phi} implies that every stable lattice is represented by infinitely many energy wells in $\calM$. Minimisers of $\tilde{\phi}(\C;T)$ at high and low temperatures correspond to the structure in austenite and martensite, respectively. For instance, the EPN of the square lattice in a square to oblique transition contains the well for the square parent phase, $S_0$, and four wells of oblique ($O_1-O_4$); these wells are infinitely replicated in $\calM$ outside the $S_0$ EPN (see Fig.~\ref{fig:Mspace_Projection_CopiesFD}). 

In the study of martensitic transformations, it is convenient to take the lattice in austenite as a reference and measure the change of energy associated with deformations from such structure. Let $\u_a$ and $\v_a=F_a^b \u_b$ be the lattice vectors of the austenite structure and a deformed lattice, respectively. The energy of the deformed lattice can be written as a function of the stretch tensor~\cite{PitteriZanzotto2003},  $\E=(\F^T\F-\Id)/2$, as follows:
\begin{equation}
\phi(\E;T)=\tilde{\phi}(u_a^T(2\E+\Id)u_b;T)-\tilde{\phi}(u_a^Tu_b;T)~,
\end{equation}
where $\Id$ is the identity matrix. The reference lattice corresponds to $\F=\Id$ or, equivalently to $\E=\mathbf{0}$, and has energy $\phi(\mathbf{0})=0$.

For nets, $\E$ is a $2\times2$ symmetric matrix containing at most three independent elements. Any transformation path in $\calM$ can therefore be parametrised by three scalar quantities $e$, $e'$ and $e''$. Transitions between different energy wells in $\calM$ are associated with non-convex dependence of the energy on one or more of these scalar parameters. Here we illustrate the effects of the energy periodicity following the assumption proposed in Refs.~\cite{PerezReche2007PRL,PerezReche_CMT2009} that $\phi(e,e',e'')$ is a non-convex function of one of the parameters, $e$, and it is convex with respect to $e'$ and $e''$. More explicitly, we use the following expression for the energy:
\begin{equation}
\label{eq:phi_ee'e''}
\phi(e,e',e'')=f(e,h';T)+\frac{C'}{2}(e')^2+\frac{C''}{2}(e'')^2~,
\end{equation}
where $C'$ and $C''$ are elastic constants and $f(e,h';T)$ is a non-convex function of $e$; the variable $h'$ is a stress coupled to $e$.

\begin{figure}[h]
\begin{center}
\includegraphics[width=10cm]{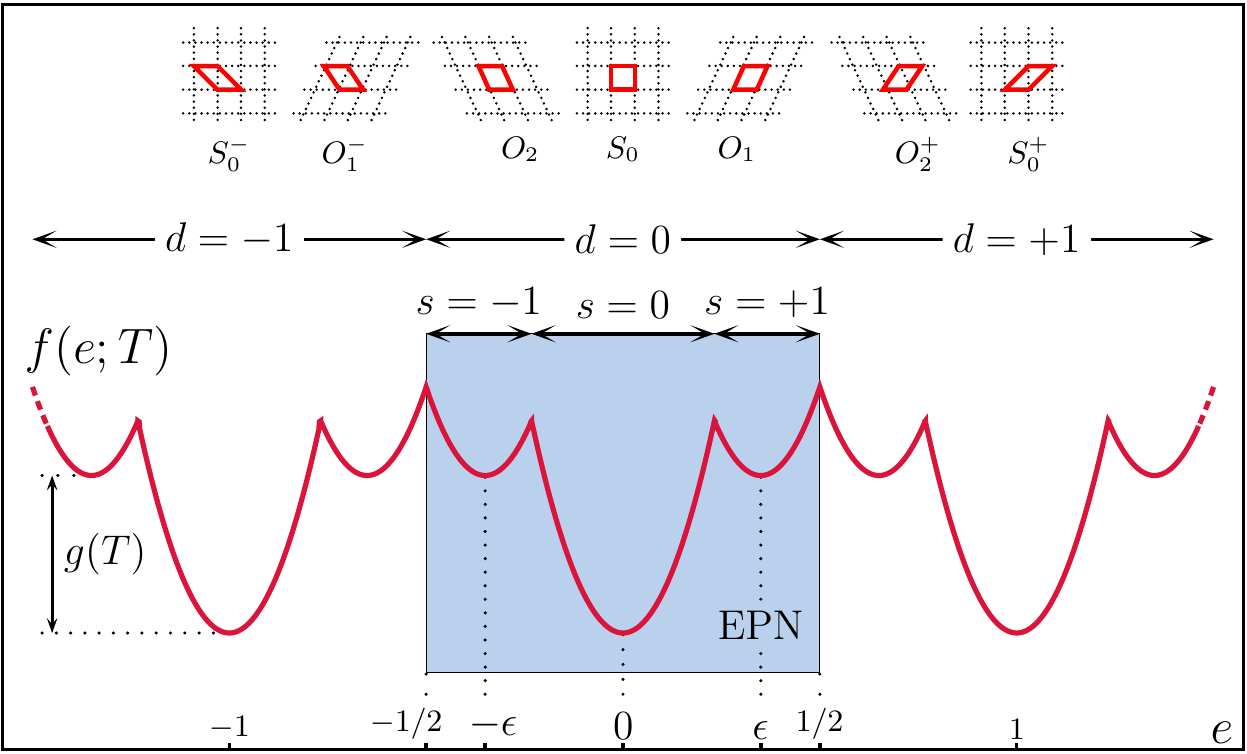}
\end{center}
\caption{Piece-wise parabolic approximation of the energy of a solid undergoing a square-oblique martensitic phase transition through a direct shear $e$ (e.g. follows a path in $\calM$ indicated by the dashed line in Fig.~\ref{fig:Mspace_Projection_CopiesFD}). The shaded region shows the EPN of the $S_0$ square net which is used as a reference frame. The insets on top show the shape of the element corresponding to the bottoms of the wells. The dashed lines indicate the lattice structure; the shapes of the mesoscopic elements are shown in red. The shapes in the domain with $d=0$ correspond to elements in square austenite, $S_0$, and two variants of martensite, $O_1$ and $O_2$. The rest of shapes are slipped lattices identical to those of the basic shapes (compare the dashed lines). [Adapted from~\cite{PerezReche2007PRL}, Fig. 1, pg. 075501-2]. }
\label{fig:Multiwell_Parabolic} 
\end{figure}

For a square-oblique transition as the one shown by the dashed line in Fig.~\ref{fig:Mspace_Projection_CopiesFD}, we approximate $f(e,h';T)$ by a piece-wise parabolic function of the direct shear strain parameter, $e=\beta/a$ (cf. Eq.~\eqref{eq:F_square-oblique}):
\begin{align}
f(e,h';T)=
\sum_{s=-1}^1 \sum_{d \in \Z} & \left[\frac{1}{2}\left(e-w(s,d)\right)^2+g(T)s^2 - h' e \right] \nonumber\\
&\times \Theta(e-e^-(s,d)) \Theta(e^+(s,d)-e)~.
\label{eq:f_homogeneous}
\end{align}
Fig.~\ref{fig:Multiwell_Parabolic} shows a plot of $f(e,h';T)$ as a funtion of $e$ at fixed temperature. 
In Eq.~\eqref{eq:f_homogeneous}, $\Theta(x)$ is the Heaviside step function which is zero for $x<0$ and one for $x \geq 0$. The domain of periodicity of the energy is given by the integer variable $d$ which takes the value $d=0$ for the EPN of $S_0$. For given $d$, the variable $s$ may take the value 0 or $\pm 1$ for austenite and the two variants of oblique martensite, respectively. The bottoms of the wells are located at  $e = w(s,d) =d+\epsilon s$, where $\epsilon$ is the transformation strain. Weak transformations correspond to values of $\epsilon<1/2$; reconstructive transformations correspond to the limiting value $\epsilon=1/2$. The energy difference between martensite and austenite wells is $g(T)$ which acts as a driving parameter for thermally driven phase transitions, analogous to the driving parameter in the RFBEG described in Sect.~\ref{SubSec:Spins-Martensites}. The functions $e^{\pm}$ give the limits of stability of each well which depend on $g(T)$ and the well identity variables $s$ and $d$ as follows:
\begin{equation}
\label{e_c}
e^{\pm}(s,d)=%
\begin{cases}
\epsilon s+d \pm (\epsilon/2+g/\epsilon), \;\; & \mbox{AM phase change,  $s=0 \rightarrow s'=\pm 1$},\\
\epsilon s+d \pm (\epsilon/2-g/\epsilon), \;\; & \mbox{MA phase change, $s=\mp 1 \rightarrow s'=0$},\\
\epsilon s+d \pm (1-2\epsilon)/2, \;\; &\mbox{Slip, $(s,d)=(\pm 1,d) \rightarrow (s',d')=(\mp 1,d \pm 1)$}.
\end{cases}
\end{equation}
Here, AM and MA refer to Austenite-Martensite and Martensite-Austenite phase changes, respectively.

\section{Heterogeneous deformations. Mesoscopic description}
\label{Sec:Hetero_Deformation}

In order to account for the complex microstructure of martensites, the theory described in Sect.~\ref{Sec:Homogeneous-Deformation} must be extended to incorporate heterogeneous deformations. Mesoscopic models~\cite{Shenoy_BookBenasque2005,PerezReche2007PRL,PerezReche_CMT2009} exploit the fact that the deformation in the martensitic phase is in fact homogeneous within domains of \emph{linear dimensions} which range from several nanometres to millimetres. In 2D crystals, such domains represent homogeneous regions at mesoscopic scales between atomistic ($\sim \r{A}$) and macroscopic ($\sim$cm) distances.  Following this observation, mesoscopic models capture the heterogeneity of deformations by assuming a set of elastically compatible elements which mechanically behave as multi-stable snap-springs~\cite{PerezReche2007PRL,PerezReche_CMT2009}. Each snap-spring represents a homogeneous deformation at mesoscopic scales but an ensemble of snap-springs can describe heterogeneous deformations at macroscopic scales. Similar approaches have been applied to study the plastic deformation of crystals~\cite{Salman-Truskinovsky_PRL2011,Salman-Truskinovsky_IntJEngSci2012} and amorphous solids~\cite{Rodney2011a}. For crystals with typical lattice spacings around $5 \AA$, the linear dimension of martensite domains corresponds to $10 - 10^6$ lattice units. Accordingly, a snap-spring can be well approximated by an infinite lattice in most cases so that its deformation obeys the rules given in Sect.~\ref{Sec:Homogeneous-Deformation}. In the limit of very small snap-springs compared to the crystal, this assumption becomes equivalent to the Cauchy-Born rule used to derive continuum theories of lattices~\cite{PitteriZanzotto2003,Bhattacharya2003}. In such theories, snap-springs become points $\bfr$ in the continuum space occupied by the solid. The Cauchy-Born rule assumes that, under a deformation field $\F(\bfr)$, the lattice vectors at $\bfr$ transform as $\v_a(\bfr)=F_a^b(\bfr) \u_b(\bfr)$, i.e. transform as a homogeneous infinite lattice located at a point $\bfr$.

\begin{figure}[h]
\begin{center}
\includegraphics[width=12cm]{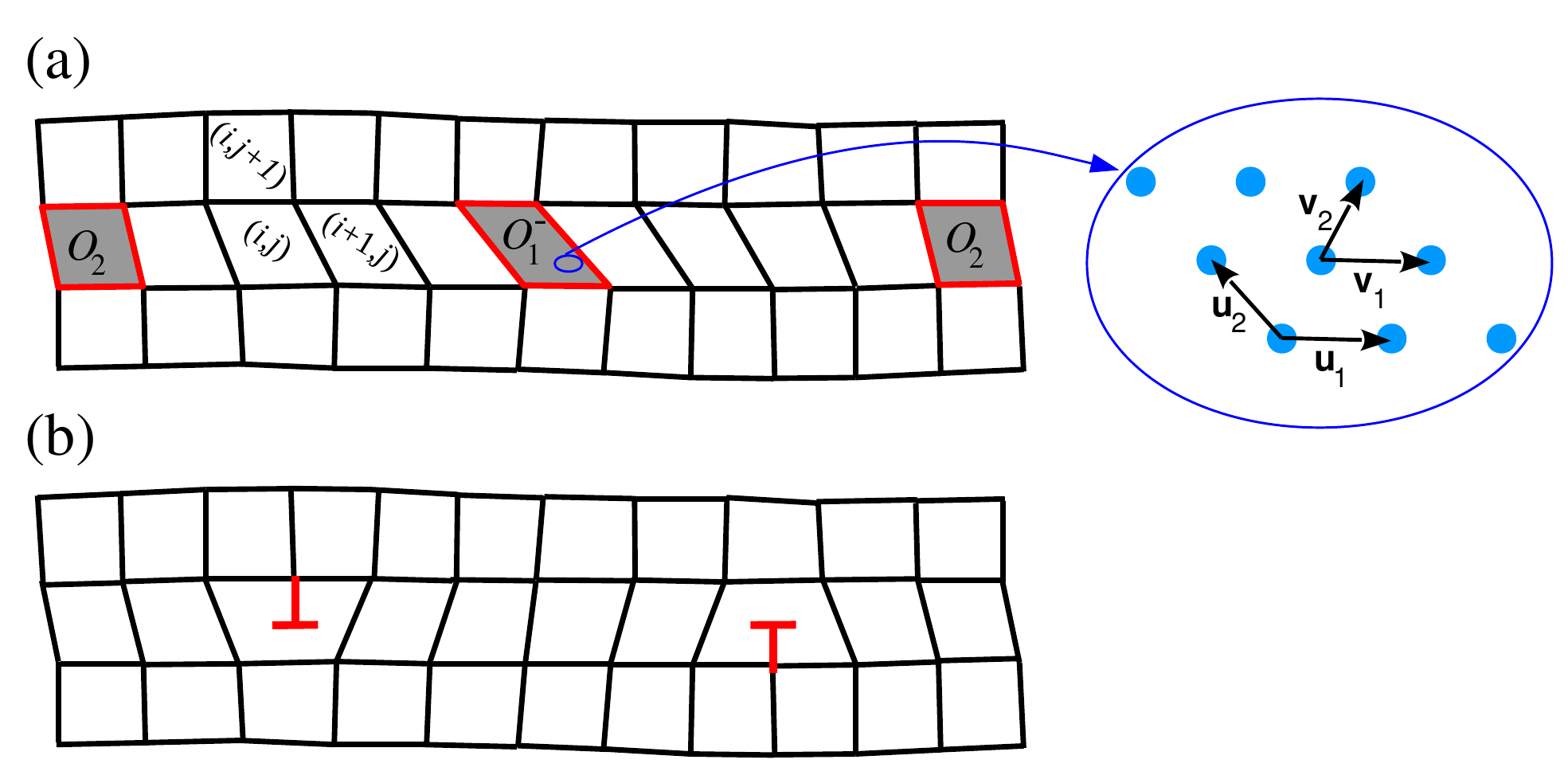}
\end{center}
\caption{(a) Description of the deformation of a 2D solid in terms of mesoscopic kinematically compatible elements (snap-springs) placed on a square grid. The lattice structure of the snap-springs marked by $O_2$ is oblique and belongs to the EPN of the square lattice $S_0$ (see Fig.~\ref{fig:Mspace_Projection_CopiesFD}). The snap-spring in the well $O_1^-$ also corresponds to an oblique lattice but it is outside the EPN of $S_0$.  Such lattice is shown in the inset where two possible bases are indicated. The base $\{\v_1,\v_2\}$ shows that the lattice is just a copy of the lattice $O_1$ outside the $S_0$ EPN. (b) Identical deformation field as in (a) showing that a mixture of snap-springs belonging to different EPNs (in this case $O_2$ and $O_1^-$) can be viewed as a solid with mesoscopic dislocations.}
\label{fig:Snap-springs} 
\end{figure}

Elastic compatibility between adjacent snap-springs means that they must fit together perfectly. This imposes constraints on the discrete deformation field, 
$\F(\bfr)$, 
defined by snap-springs~\cite{Braun}. Here, $\bfr=(i,j)$ is the coordinate of a snap-spring in the square grid of snap-springs, as indicated in Fig.~\ref{fig:Snap-springs}. For relatively small deformations (linear elasticity), the constraints are given by the St. Venant compatibility~\cite{Shenoy1999,Lookman2003,Shenoy_BookBenasque2005}, 
\begin{equation}
\label{eq:StVenant}
\De \times (\De\times \E^l(\bfr))^T=\mathbf{0}~,
\end{equation}
where $\E^l(\bfr)$ is the stress tensor which approximates the stretch tensor in linear elasticity. The symbol $\De$ is a discrete difference operator analogous to the gradient vector in continuum systems. Given a field $A(\bfr)$, it acts as $\De A(\bfr)=(A(i+1,j)-A(i,j),A(i,j+1)-A(i,j))$. 

The snap-spring model provides a unified description of phase transitions (between wells within an EPN) and slip (transitions between energy wells belonging to different EPNs). Note that a compatible deformation field $\F(\bfr)$ can involve snap-springs that do not belong to the same EPN (cf. snap-springs in wells $O_2$ and $O_1^-$ in Fig.~\ref{fig:Snap-springs}). Therefore, slip does not necessarily lead to a violation of elastic compatibility between snap-springs.

An interesting consequence of elastic compatibility in heterogeneous deformation fields is that slip can occur locally~\cite{Balandraud_Zanzotto2007} even when the phase transition of isolated snap-springs is weak (i.e. when martensite wells belong to the EPN of austenite). Recall that the theory described in Sect.~\ref{SubSec:Weak-Reconstructive} would only predict the occurrence of slip for reconstructive transitions. In contrast, local slip is indeed possible for heterogeneous \emph{weak} deformations due to elastic compatibility which brings an interaction between snap-springs that may force some of them to explore regions of $\calM$ outside the EPN of the parent phase. We then conclude that slip is much more frequent in heterogeneous deformations than it is for homogeneous crystals. In addition, the EPN concept is not enough to determine the conditions for slip in heterogeneous deformation fields. 

\begin{figure}[h]
\includegraphics[width=10cm]{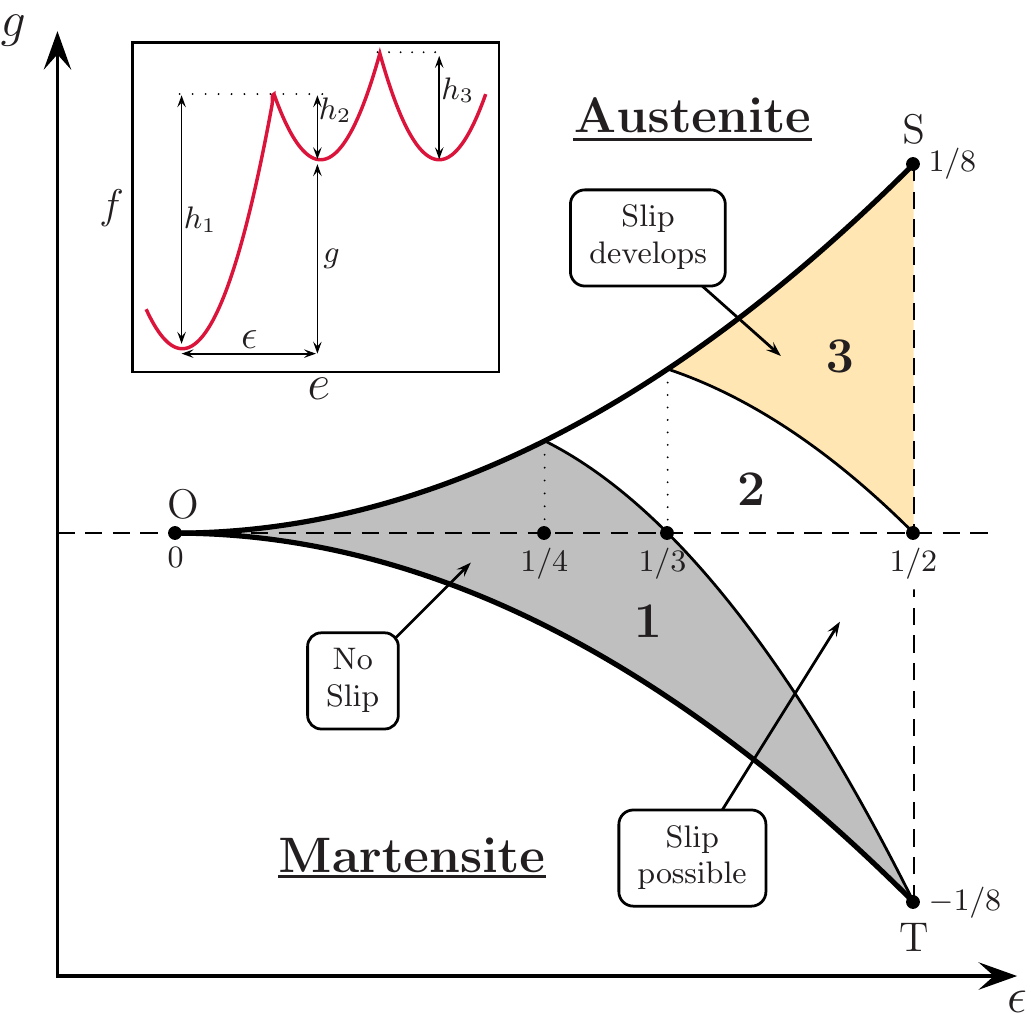}
\caption{Slip phase diagram for a snap-spring in the space $(\epsilon,g(T))$. The inset indicates the energy
  barriers $h_1$, $h_2$ and $h_3$ used to define three characteristic
  zones in the domain of phase coexistence \emph{SOT} [Adapted from \cite{PerezReche_CMT2009}, Fig. 5].}
\label{fig:Phase_Diagram_Homogeneous} 
\end{figure}

Fig.~\ref{fig:Phase_Diagram_Homogeneous} shows a phase diagram presented in Ref.~\cite{PerezReche_CMT2009} that quantifies the possible slip within a snap-spring due to interaction with other snap-springs in the crystal. The stability boundaries for the
martensite and the austenite phases can be derived from the limits given by Eq.~\eqref{e_c} and are indicated by the lines
\emph{OS} and \emph{OT}, respectively. The possibility of slip in weak transformations can be heuristically analysed by comparing the energy barriers indicated in the inset of the figure for each transition type: austenite-martensite ($h_1$), martensite-austenite ($h_2$) and slip ($h_3$). One can distinguish three regions in the space $(\epsilon,g(T))$. In region 1, $h_3>h_1$ and slip can be typically neglected. In region 2, $h_2<h_3<h_1$ and slip is possible but not likely. Finally, in region 3 the barriers satisfy $h_3<h_2<h_1$ meaning that the barrier for slip is smaller than the rest and slip may easily develop. Note that region 3 is close to the limit $\epsilon=1/2$ for reconstructive transitions but plasticity can already develop for $\epsilon<1/2$.

Following the arguments presented in Sect.~\ref{SubSec:Energy-Homogeneous}, the deformation of a snap-spring $i$ can be given by three scalar parameters $e_i$, $e_i'$ and $e_i''$. The energy of a solid consisting of $N$ snap-springs can then be expressed as a sum over the energies of individual snap-springs as follows:
\begin{equation}
\label{eq:phi_ee'e''_snapspring}
\Phi_{\text{solid}}(\e,\e',\e'')=\sum_{i=1}^N \phi(e_i,e_i',e_i'')~,
\end{equation}
where $\e=\{e_i;\;i=1,2,\dots,N\}$; the sets $\e'$ and $\e''$ are defined analogously. Here, one should also keep in mind that the parameters $\e$, $\e'$ and $\e''$ are not independent from each other due to elastic compatibility.

\section{The random snap-spring model (RSSM)}
\label{Sec:RSSM}

Mesoscopic models of the type described in the previous section were proposed in Refs.~\cite{PerezReche2007PRL,PerezReche_CMT2009,Perez-Reche_PRB2016} and Ref.~\cite{PerezReche_PRL2008} for thermally and mechanically driven materials, respectively. In this section, we present an extended formulation of these models to encompass the two types of driving. 

\begin{figure}[b]
\includegraphics[width=8.0cm]{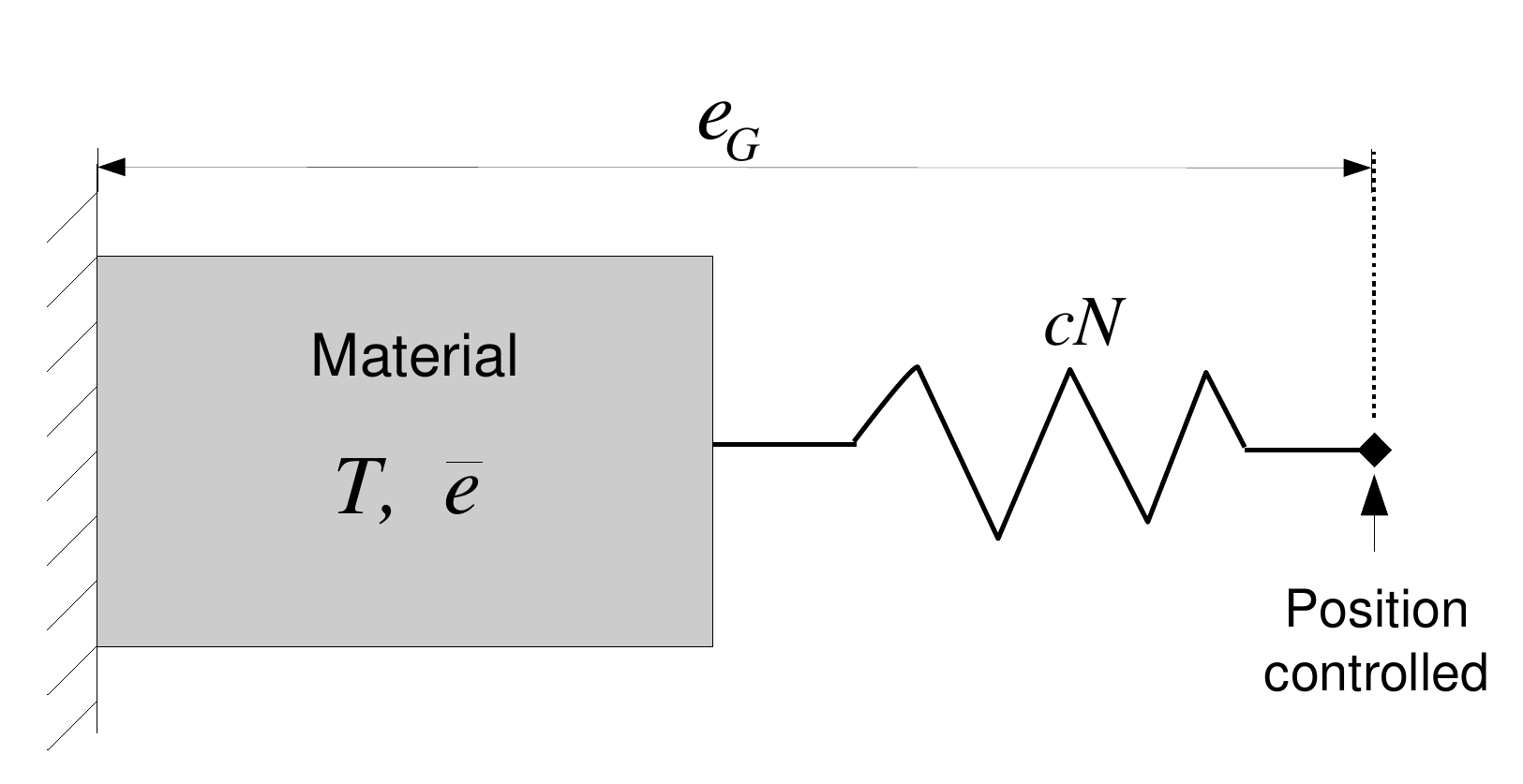}
\caption{Schematic representation of a solid attached to an elastic loading device with stiffness $c\,N$. Structural phase changes can be induced by sweeping the temperature, $T$, or the global strain of the system, $e_G$, by controlling the position of the loading device.}
\label{fig:Loading_Device_3D} 
\end{figure}

The energy $\Phi$ of the system consists of two contributions, $\Phi_{\text{solid}}$ and $\Phi_{\text{mech}}$, associated with the crystal (see Eq.~\eqref{eq:phi_ee'e''_snapspring}) and the mechanical load, respectively. 
\begin{equation}
\label{eq:Phi-SolidMech}
\Phi=\Phi_{\text{solid}}+\Phi_{\text{mech}}~.
\end{equation}

As shown in the scheme of Fig.\ref{fig:Loading_Device_3D}, mechanical driving is applied through an elastic device attached to the system. The driving parameter is the global ``elongation'' of the system, $e_{G}$, which is assumed to be parallel to the order parameter $e$ for simplicity. The energy associated with the loading device is  
\begin{equation}
\label{eq:Phi-Mech}
\Phi_{\text{mech}}=\frac{cN}{2}(e_G-\bar{e})^2~,
\end{equation}
where $\bar{e}=\frac{1}{N}\sum_{i=1}^N e_i$  is the
average strain of the system of snap-springs. The stiffness of the loading device is $c\, N \geq 0$, where $c$ is an specific stiffness which is assumed to be independent of the system size. The energy $\Phi_{\text{solid}}$ increases with the system size (i.e. it is extensive). By defining the stiffness $c\, N$ as an extensive variable we make sure that the energy $\Phi_{\text{mech}}$ is also extensive and can represent a finite contribution to the total energy irrespective of the system size $N$. In addition, chosing an extensive stiffness ensures that finite elongations of the loading device, $e_G-\bar{e}$, can represent a non-zero driving force for the snap-spring system irrespective of $N$. The value of the specific stiffness, $c$, can vary between $0$ and $\infty$. The limit as $c=\infty$ defines a hard device where the loading parameter must be $e_G=\bar{e}$ in order for $\Phi_{\text{mech}}$ and thus $\Phi$ to be finite. In this case, controlling $e_G$ is equivalent to controlling the average strain of the system. In the opposite limit as $c \rightarrow 0$, the driving force vanishes unless $e \rightarrow \infty$ and the product $ce_G$ remains finite. This is a situation in which the system is loaded by a very soft spring with large elongation and controlling $e$ becomes equivalent to applying a stress $\sigma \equiv ce_G$.

The energy of the solid is assumed to be given by Eq.~\eqref{eq:phi_ee'e''_snapspring} with $\phi(e_i,e_i',e_i'')$ taking the form proposed in Eq.~\eqref{eq:phi_ee'e''}, i.e.
\begin{equation}
\label{eq:Phi-Solid}
\Phi_{\text{solid}}(\e,\e',\e'')=\sum_{i=1}^N \left[f(e_i,h_i';T)+\frac{C'}{2}(e_i')^2+\frac{C''}{2}(e_i'')^2 \right]~,
\end{equation}
where $f_i(e_i,h_i';T)$ is given by Eq.~\eqref{eq:f_homogeneous}. The local stress variables, $\{h_i'\}$, are defined as quenched random variables analogous to the random fields in random-field models (see Sect.~\ref{Sec:Spin-Models}). Within the context of the RSSM, random fields provide an effective description of the effects of, e.g. local impurities or dislocations at atomic scales (crystal dislocations). 
Point defects are responsible for interesting effects in martensites such as tweed precursors or strain-glass behaviour~\cite{Kakeshita-Planes_Book2012}. Such phenomena have been modelled for weak transformations using Landau theories~\cite{Kartha1991,Kartha1995,Lloveras_PRL2008,Wang-Ren_PRL2010} and spin models~\cite{Sherrington2008,Vasseur-Lookman_PRB2010}. By including quenched random fields in the RSSM, we extend such theories to account for evolving disorder associated with transformation-induced slip and its interplay with quenched disorder (see Sect.~\ref{Sec:ThermalDriven}). 
Crystal dislocations are also an important factor in martensites.  
Arbitrary distributions of crystal dislocations within snap-springs may in general lead to a non-zero Burgers vector at mesoscopic scales and this translates into a lack of elastic compatibility between snap-springs. The non-compatibility effects were analysed in~\cite{Groger_PRB2008,Groger_PRB2010} using Landau theories for weak phase transitions. In principle, such methods could be extended to be included in the model described here. We will however focus on systems with particular distributions of crystal dislocations giving a zero net Burgers vector within snap-springs so that snap-springs are elastically compatible.

The system is either thermally or mechanically driven through variation of the parameter $g(T)$ or the global elongation $e_G$, respectively. Under quasistatic driving and negligible thermal fluctuations~\cite{PerezReche2001,Salje_PRB2011}, the model exhibits avalanche dynamics which can be described in terms of automata~\cite{PerezReche2007PRL,PerezReche_CMT2009}. Avalanche dynamics are characterised by quiescent periods in which the system remains in a local energy minimum with fixed configuration fields $\s=\{s_{i,j}\}$ and $\d=\{d_{i,j}\}$ interrupted by avalanches leading to changes in $\s$ and/or $\d$. Avalanches start when the stability condition $e^- < e_i < e^+$ is violated by at least one snap-spring $i$ and stop when the condition is again satisfied by all snap-springs. We now describe the steps followed to obtain an automaton representation of the model.

The relaxation of the harmonic variables $\e'$ and $\e''$ is assumed to be instantaneous with respect to the time scales of both driving and relaxation of the primary order parameter, $\e$. In this case, $\e'$ and $\e''$ can be adiabatically eliminated by setting $\partial \Phi/\partial e_i'=0$ and $\partial \Phi/\partial e_i''=0$ subject to the elastic compatibility conditions~\cite{Shenoy1999,Lookman2003,Shenoy_BookBenasque2005}. This allows the energy of the system to be expressed in terms of the primary order parameter only~\cite{PerezReche2007PRL,PerezReche_CMT2009,Perez-Reche_PRB2016}:
\begin{equation}
\label{eq:energy_e}
\tilde{\Phi}(\e)=\frac{1}{2} \sum_{i,j=1}^{N} K_{ij} e_i
e_j+ \sum_{i=1}^N f_i(e_i,h_i';T)+\frac{cN}{2}(e_G-\bar{e})^2~.
\end{equation}
From this expression,  the RSSM can be viewed as a set of snap-springs with energy $f_i(e_i,h_i';T)$ which, due to elastic compatibility, interact with an interaction kernel ${\bf K}=\{K_{ij}\}$. 

The variables $\e$ are assumed to obey overdamped dynamics given by the following set of equations:
\begin{equation}
\label{eq:dynamics_RSSM}
\frac{1}{\gamma} \frac{\partial e_i}{\partial t} = - \frac{\partial
  \tilde{\Phi}}{\partial e_i}, \hspace{20pt} i=1,2,\dots,N~,
\end{equation}
where $\gamma$ is the ratio between the rate of relaxation of the
system to the local minimum of energy and the rate of driving.
 In the quasistatic limit, $\gamma \rightarrow \infty$, and the left
hand side in Eqs.~(\ref{eq:dynamics_RSSM}) vanishes. As a result, the
dynamics project onto the local minima of $\tilde{\Phi}$
which form a discrete set of branches with state variables
$\e(e_G;\s,\d)$. Each branch corresponds to a different configuration of
 $\s$ and $\d$. Minimization of $\tilde{\Phi}$ gives the
following expression for the strain along equilibrium branches:
\begin{equation}
\label{eq:Equilibrium_General}
e_i=\frac{ck_i}{1+ck_{\infty}}e_G+ 
\epsilon \sum_{j=1}^N \left( J_{ij}-\frac{k_{ij}(c)}{N}\right)s_j+\sum_{j=1}^N \left( J_{ij}-\frac{k_{ij}(c)}{N}\right)d_j+h_i~.
\end{equation}
Here, $\{J_{ij}\}$ are the elements of a matrix $\J=(\Id +
\K)^{-1}$, $h_i=\sum_jJ_{ij} h_j'$ are renormalised disorder variables, $k_i=\sum_{j} J_{ij}$, 
$k_{\infty}=\frac{1}{N} \sum_{i}k_i$,  and $\{k_{ij}(c)\}$ are the
elements of an effective stiffness matrix defined as
\begin{equation}
\label{eq.stiff_mat_ij}
k_{ij}(c)=\frac{ck_ik_j}{1+ck_{\infty}}~.
\end{equation}
The values taken by $k_{ij}(c)$ are limited to the interval $0 \leq
k_{ij} < k_ik_j/k_{\infty}$ where the lower and upper bounds correspond
to $c=0$ and $c \rightarrow \infty$, respectively.

\subsection{The RSSM as a random-field model}

Comparison of Eq.~\eqref{eq:Equilibrium_General} with the local field for the RFIM (Eq.~\eqref{eq:RFIM_LocalField}) reveals that, along the equilibrium branches, the RSSM behaves as a random-field model with applied field proportional to $e_G$, and two types of random fields. The first type, $h_i$, represents generic heterogeneity while the second type,  
\begin{equation}
\label{eq:h_slip}
h_i^{\text{s}}=\sum_{j=1}^N \left( J_{ij}-\frac{k_{ij}(c)}{N}\right)d_j~
\end{equation}
is associated with slip. We will assume that random fields, $h_i$, are quenched and are given by a Gaussian distribution with zero mean and standard deviation $r$. In contrast, slip fields, $h_i^{\text{s}}$ are not quenched in general since they are associated with the slip variables, $\d$, which evolve as the system is driven through the phase transition and slip is generated. 

The behaviour of the RSSM along equilibrium branches reduces
  to that of a random-field model for the discrete variables $s_i$
  and $d_i$.  The interaction between these variables consists of two contributions: the kernel $\J=\{J_{ij}\}$ originated by elastic compatibility and an infinite-range contribution $k_{ij}(c)/N$ associated with the stiffness of the mechanical constraint.  

Substituting the strain given by Eq.~\eqref{eq:Equilibrium_General} in Eq.~\eqref{eq:energy_e} and using Eq.~\eqref{eq:f_homogeneous} with  $w_i=\epsilon s_i+d_i$ on gets the following energy for the RSSM along equilibrium branches:
\begin{equation}
\hat{\Phi}= \Phi_{\text{ss}}+\Phi_{\text{dd}}+\Phi_{\text{sd}}+\Phi_{\text{h}}+\Phi_{\text{mech}}~.
\end{equation}
Here, we have neglected constant contributions and have defined
\begin{align}
\label{eq:hatPhi_ss}
&\hat{\Phi}_{\text{ss}} = -\frac{\epsilon^2}{2} \sum_{i,j}\left(J_{ij}-\frac{k_{ij}(c)}{N}
  -2\frac{g(T)}{\epsilon^2} \delta_{ij}\right)s_i s_j~,\\
\label{eq:hatPhi_dd}
&\hat{\Phi}_{\text{dd}} = -\frac{1}{2} \sum_{i,j} \left(J_{ij}-\frac{k_{ij}(c)}{N} \right) d_i d_j~,\\
\label{eq:hatPhi_sd}
&\hat{\Phi}_{\text{sd}} = -\epsilon \sum_{i,j} \left(J_{ij}-\frac{k_{ij}(c)}{N} \right) s_i d_j,\\
\label{eq:hatPhi_h}
&\hat{\Phi}_{\text{h}} =-\sum_i h_i(\epsilon s_i+d_i)~,\\
\label{eq:hatPhi_mech}
&\hat{\Phi}_{\text{mech}} = \frac{cN e_G}{1+c k_{\infty}} \left(\frac{e_G}{2}
  -\frac{k_{\infty}}{N} \sum_{i} (\epsilon s_i+d_i)\right)~.
\end{align}
The contribution $\hat{\Phi}_{\text{ss}}$ gives the interaction between snap-springs associated with their phase state, $\s$. $\hat{\Phi}_{\text{dd}}$ gives an analogous interaction for the slip state, $\d$. The energy $\hat{\Phi}_{\text{sd}}$ gives the interaction between phase transition and slip. $\hat{\Phi}_{\text{h}}$ accounts for the effects of snap-spring heterogeneity on the phase and slip states of snap-springs. Finally, $\hat{\Phi}_{\text{mech}}$ is the energy associated with the mechanical loading. 

\section{Thermally-driven transformations}
\label{Sec:ThermalDriven}

In order to illustrate the behaviour of thermally-driven structural transformations, we study the RSSM for mechanically unconstrained 2D solids by setting $c=0$. The interaction kernel, $\J$, can in principle be explicitly calculated through minimisation of the energy of the system with respect to the strain fields $\e$, $\e'$, $\e''$, as described in Sect.~\ref{Sec:RSSM}. An explicit calculation of this type is presented in Ref.~\cite{Perez-Reche_PRB2016} for a highly anisotropic model which makes the calculation more tractable. A simpler alternative method consists in assuming periodic boundary conditions and then calculate the Fourier transform of the kernel $\K$ in Eq.~\eqref{eq:energy_e} from which one can obtain $\J=(\K+\Id)^{-1}$~\cite{Kartha1995,Shenoy1999,Lookman2003,Salman-Truskinovsky_PRL2011,Salman-Truskinovsky_IntJEngSci2012}. Irrespective of the method used to calculate the kernel $\J$, one always finds the following key features: anysotropy, sign indefiniteness and long-range decay. For illustration purposes, here we assume a short-range anisotropic and non-positive definite $\J$ as in~\cite{PerezReche2007PRL,PerezReche_CMT2009}; more realistic long-range interactions have been considered in \cite{Perez-Reche_PRB2016} within a similar setting. We use the following kernel:
\begin{equation}
\label{Jij}
J_{ij}=
\begin{cases}
J_0, & i=j\\
J_1>0, & i \mbox{ n.n. } j \\
-J_2<0, & i \mbox{ n.n.n. } j \\
0, & \mbox{otherwise},
\end{cases}
\end{equation}
where `n.n.' and `n.n.n.' indicate nearest and next-to-nearest
neighbors, respectively. We further assume that $\sum_{i,j}J_{ij}=0$ which ensures that the inhomogeneity of the field ${\bf w}=\epsilon \s+\d$ is penalised. For snap-springs on a square grid, this condition gives $J_0=-4(J_1-J_2)$. 

In the following two subsections, we present results obtained through numerical simulation of a $501\times 501$ ensemble of snap-springs placed on a square grid with open boundary conditions. The interaction terms are set to $J_1=0.062$ and $J_2=0.03$. The system is initially prepared in a homogeneous austenite phase with a minimal dislocation loop in the centre. Thermal loading is applied by cyclically sweeping $g(T)$ through the complete transformation. 

\subsection{Evolution of slip in systems without quenched disorder}
\label{SubSec:QuenchedDisorder}

We now study the effect of transformation-induced defects in systems in which the only source of heterogeneity between snap-springs is associated with the slip field $\d$; Gaussian quenched disorder is set to $r=0$.

\begin{figure}
\centering
{\includegraphics[width=12.0cm]{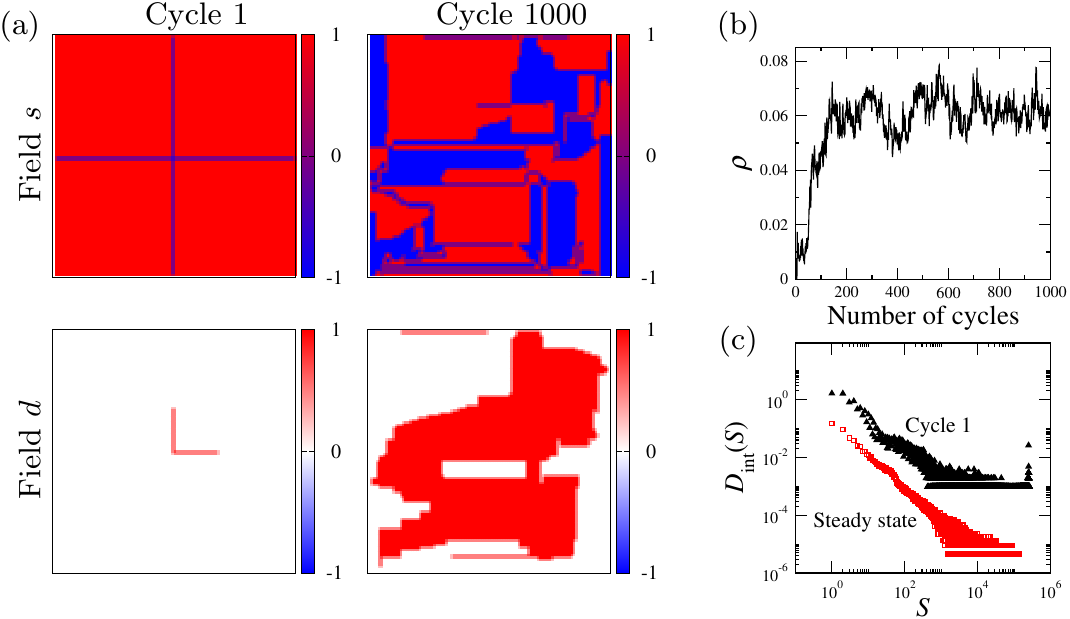}}
\caption{ \label{Microstructure_Tdriv_NoDisorder} Evolution under thermal cycling of a solid without  quenched disorder undergoing a close to reconstructive transition with $\epsilon=0.47$. (a) The upper and lower panels show the phase ($\s$) and slip ($\d$) 
  microstructures, respectively, after cycle 1 and cycle 1000. A
  dislocation loop was initially placed in the centre of the system.
 (b) Dislocation density $\rho$ during the first 1000
  cycles. (c) Distribution of the avalanche sizes, $D_{\text{int}}(S)$, after cycle 1
  (triangles) and in the stationary state after 1000 cycles
  (squares).}
\end{figure}

Fig.~\ref{Microstructure_Tdriv_NoDisorder} shows the results for a system with $\epsilon=0.47$ inside region 3 in Fig.~\ref{fig:Phase_Diagram_Homogeneous} where transformation-induced slip is expected. The upper panels in Fig.~\ref{Microstructure_Tdriv_NoDisorder}(a) show the spatial distribution
of $\s$ in the martensitic phase after cycle 1 and after cycle 1000.
The complexity of the phase microstructure clearly increases during
the training period. In the lower panels one can see that the system develops some slip ($d \neq 0$) induced by the phase transition. The increase of slip with thermal cycling is clear in Fig.~\ref{Microstructure_Tdriv_NoDisorder}(b) which shows the density $\rho$ of nearest neighbor snap-prings 
with differing values of $d_i$. As argued in Sect.~\ref{Sec:Hetero_Deformation} and illustrated in Fig.~\ref{fig:Snap-springs}, neighbouring snap-springs with different values of $d_i$ imply dislocations at a mesoscopic scale. Accordingly, the density $\rho$ is a measure of the dislocation density at a mesoscale.
The evolution of $\rho$ is marked by a steep initial increase (training period) which
after approximately $150$ cycles leads to a steady regime (shakedown).

Fig.~\ref{Microstructure_Tdriv_NoDisorder}(c) shows the distribution of avalanche sizes, $D_{\text{int}}(S)$, calculated by pooling avalanches observed during complete cooling runs. $D_{\text{int}}(S)$ evolves from a supercritical behaviour (peak at
large values of $S$) during the first cycles towards a power law, $D_{\text{int}}(S) \sim S^{\tau'}$, in
the steady state regime. A peak at large values of $S$ indicates the occurrence of a snap event as in the low-disorder regime of random-field models presented in Sect.~\ref{Sec:Spin-Models}. The exponent of the power-law in the steady state regime takes a value $\tau' \simeq 1.2$ which is compatible with the exponent for a 2D RFBEG~\cite{Vives1995Universality}. These results suggest that thermal cycling generates slip disorder that allows the system to cross a critical manifold associated with an OD transition of the type observed for random-field models with nucleation dynamics. In Ref.~\cite{Perez-Reche_PRB2016} we present a more detailed analysis of the origin of robust criticality in the RSSM and its link to OD critical transitions. We find a different value for the exponent $\tau'_{\text{LR}} \simeq 1.6$ when considering a more realistic long-range interaction kernel $\J$. Training effects predicted by models with short-range interactions are qualitatively similar to those observed for long-range interactions  but the universality of critical avalanches is different.

\subsection{Interplay between quenched and evolving disorder}
\label{SubSec:Quenched-Evolving-Disorder}

\begin{figure}
{\includegraphics[width=12cm]{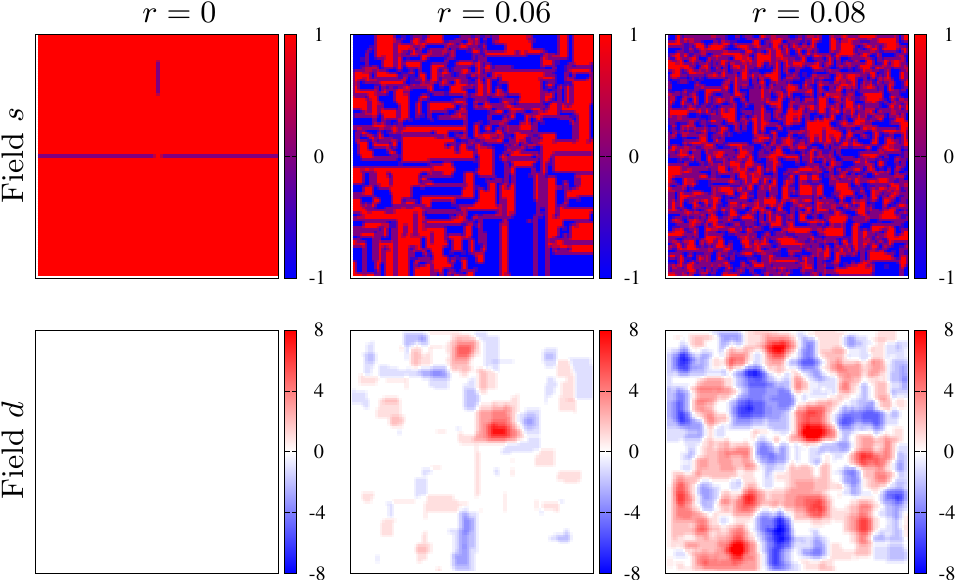}}
\caption{ \label{Microstructure_Tdriv_Disorder} Systems with quenched and evolving disorder after 1000 cycles. The transformation strain is $\epsilon=0.46$. The upper panels show the microstructure of phases, $\s$, in the martensitic phase; the lower panels show the slip field $\d$. Each column corresponds to a different degree of quenched disorder, $r=0,0.06, 0.08$.}
\end{figure}

\begin{figure}
\centering
{\includegraphics[width=11cm]{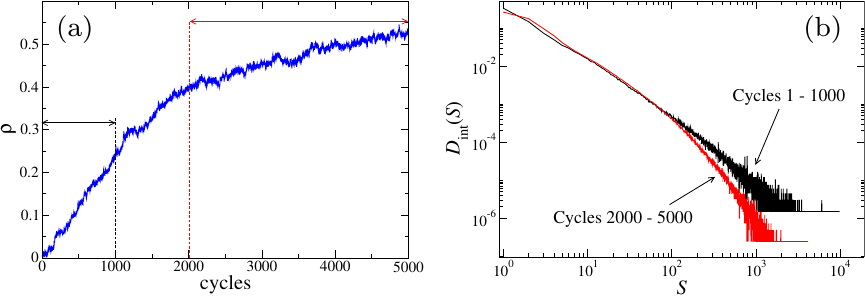}}
\caption{ \label{Effect_disorder_rho_Ds_Tdriv} (a) Dislocation density, $\rho$, during 5000 cycles in a system with quenched disorder $r=0.06$ and transformation strain $\epsilon=0.46$. (b) Avalanche size distribution pooling avalanches registered during cycles $1-1000$ and $2000-5000$.}
\end{figure}

We now explore the combined effect of Gaussian quenched disorder in intrinsically disordered solids and evolving slip disorder induced through thermal cycling. The predictions of this study are relevant to solids with impurities exhibiting, e.g., tweed precursors or strain-glass phases~\cite{Kakeshita-Planes_Book2012}. We consider a system with smaller transformation strain than in the previous section, $\epsilon=0.46$, such that slip generated in the absence of quenched disorder is negligible (see the panels for $r=0$ in Fig.~\ref{Microstructure_Tdriv_Disorder}). In general, the amount of transformation-induced slip increases for increasing degree of quenched disorder (see panels for $r=0.06$ and $r=0.08$ in Fig.~\ref{Microstructure_Tdriv_Disorder}). In other words, the RSSM predicts that a large amount of quenched microscopic defects will typically induce larger amounts of transformation-induced slip. The density of dislocations in systems with quenched disorder, $r>0$, takes larger values than for systems with $r=0$ and also takes longer to reach a steady state (compare Fig.~\ref{Effect_disorder_rho_Ds_Tdriv}(a) and Fig.~\ref{Microstructure_Tdriv_NoDisorder}(b)). The avalanche size distribution, $D_{\text{int}}(S)$, is also affected by the degree of quenched disorder. Systems with large enough $r$ develop high levels of slip under thermal cycling and obey a subcritical $D_{\text{int}}(S)$ (i.e. an exponential decay for large $S$), reminiscent of the response of random-field models in the pop regime. In addition, the cut-off of $D_{\text{int}}(S)$ at large avalanche sizes becomes increasingly pronounced as the levels of slip increase with cycling (see Fig.~\ref{Effect_disorder_rho_Ds_Tdriv}(b)).


\section{Mechanically-driven transformations}
\label{Sec:Mech-driven}

\begin{figure}
{\includegraphics[width=6.0cm]{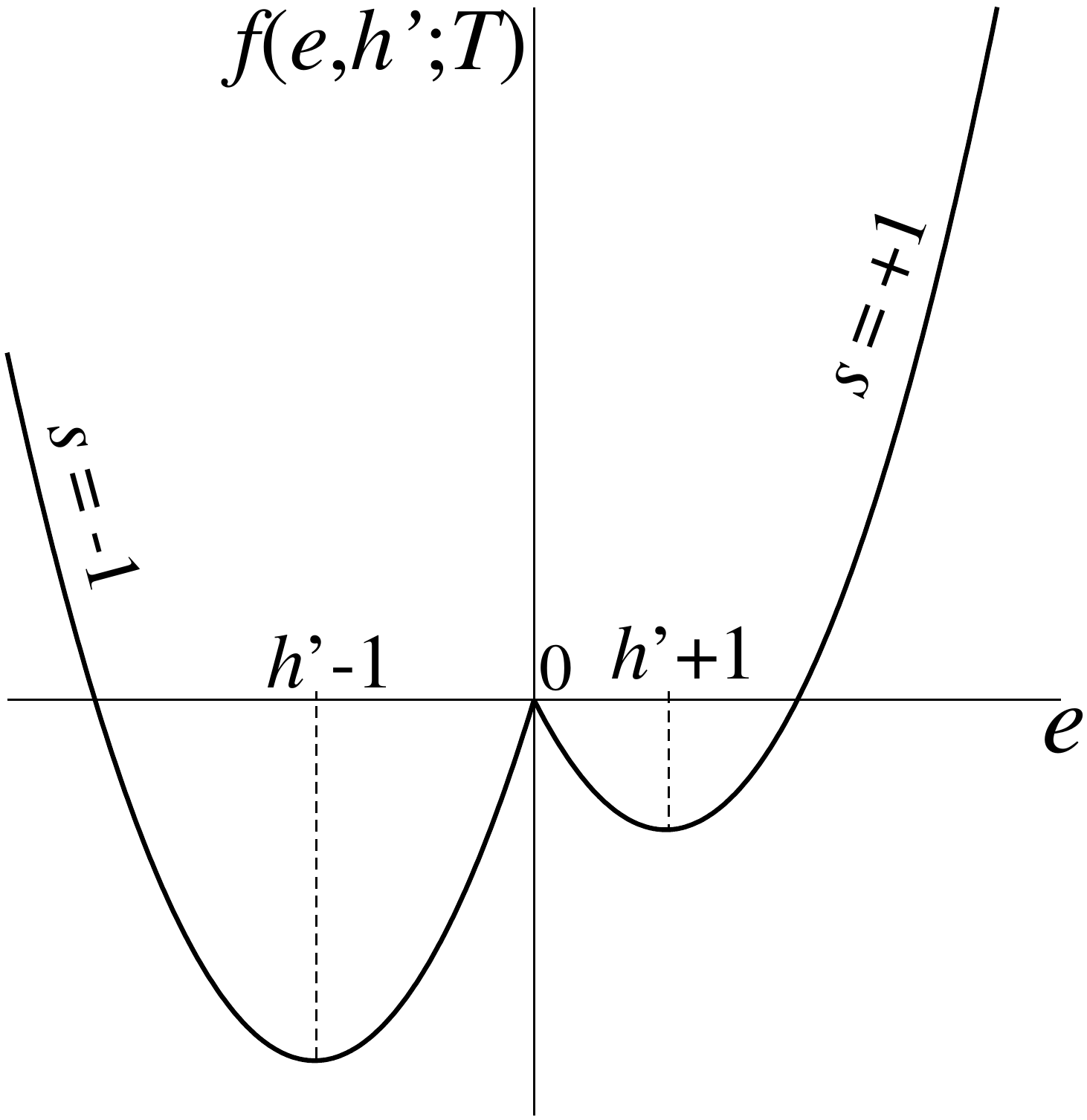}}
\caption{ \label{Double_Well} Piece-wise parabolic energy for a snap-spring at low temperature  in the limit of small transformation strain, $\epsilon$, which is taken as the strain unit. }
\end{figure}

This section presents some predictions of the RSSM for mechanically-driven systems~\cite{PerezReche_PRL2008}. We consider weak transformations with $\epsilon \rightarrow 0$ so that slip can be neglected. In this regime, it is useful to set $\epsilon$ as the unit for strain in such a way that the bottom of the martensite wells are located at $e=\pm \epsilon=\pm 1$. Let us also assume a particular case at low temperature, $g(T)<-\epsilon^2/2$, so that all the snap-springs are in the martensitic phase. The resulting energy for a snap-spring, $f(e,h';T)$ (cf. Eq.~\eqref{eq:f_homogeneous}), is illustrated in Fig.~\ref{Double_Well}. These conditions correspond to the shape-memory regime in shape-memory alloys~\cite{Christian2002v,Ortin-Planes-Delaey_ScienceHysteresis2006}. In the following subsections, we present the stress-strain curves, transformation mechanisms and avalanche statistics predicted by the RSSM depending on the degree of quenched disorder, $r$, and stiffness of the loading device, $c$.

In the numerical simulations presented below, the system is initially prepared in a state with all snap-springs in the variant $s=-1$. The global elongation, $e_G$, is then quasistatically increased until all the snap-springs have transformed to the variant $s=+1$. 
We assume relatively simple settings for the snap-spring ensemble. The snap-springs are placed on the nodes of a simple cubic lattice of linear size $L=N^{1/3}$. This will allow us to compare with the well-studied zero-temperature RFIM in 3D. The interaction kernel $\J$ takes a short-range form with $J_{ii}=J_0 \geq 0$, $J_{ij}=J_1 > 0$ between nearest neighbors, and $J_{ij}=0$ beyond nearest neighbors (in numerical simulations, we set $J_0=J_1=1$). The non-negative character of the kernel ensures that  the fraction $f_+=(\sum_i s_i/N+1)/2$ of snap-springs in the variant $s=+1$ increases monotonically with increasing $e_G$ (i.e. no backward flips occur). We use periodic boundary
conditions such that the quantity $k_i=\sum_j J_{ij}$ does not
depend on $i$ and takes the value $k_{\infty}=J_0+6J_1$ for snap-springs on a cubic lattice. The
elements of the effective stiffness matrix $k_{ij}$ defined in Eq.~\eqref{eq.stiff_mat_ij} are all equal to
\begin{equation}
\label{eq.stiffness}
k(c)=k_{\infty} \left[\frac{ck_{\infty}}{1+ck_{\infty}} \right]~.
\end{equation}
Within this simplified setting, the behaviour of the RSSM along equilibrium branches can be mapped to the RFIM described in Sect.~\ref{Sec:Spin-Models} with nearest-neighbour interaction, $\J$, and an infinite-range interaction, $J_{\text{inf}}=k(c)$.

\subsection{Stress-strain curves}
\label{SubSec:Stress-strain}

The stress is an intensive quantity defined as $\sigma=N^{-1}\frac{\partial
  \Phi}{\partial e_G}$ which, from Eqs.~\eqref{eq:Phi-SolidMech}, \eqref{eq:Phi-Mech} and \eqref{eq:Phi-Solid} reads as
\begin{equation}
\label{eq.Stress}
\sigma=c(e_G-\bar{e}).
\end{equation}
The average strain along the equilibrium branches 
can be obtained by introducing the equilibrium values $e_i$ from Eq.~(\ref{eq:Equilibrium_General}) in the definition $\bar{e}=\frac{1}{N}\sum_ie_i$. One obtains,
\begin{equation}
\label{eq.bar_e}
\bar{e}=\frac{1}{1+ck_{\infty}}\left(ck_{\infty}e_G+\frac{\sum_i k_i w_i}{N}\right).
\end{equation}
The stress corresponding to the equilibrium branches can be obtained from Eqs.~(\ref{eq.bar_e}) and (\ref{eq.Stress}) which give
\begin{equation}
\label{eq.Stress_eq}
\sigma=\frac{c}{1+ck_{\infty}}
\left[ e_G- \frac{\sum_i k_i w_i}{N}\right]~.
\end{equation}
The soft-device limit corresponds to $c \rightarrow 0$ with finite $ce_G$ which gives $k_i=0$ and $\sigma = c e_G$, as argued in Sect.~\ref{Sec:RSSM}. The limit $c \rightarrow \infty$ corresponds to a hard device with stress
\begin{equation}
\label{eq.Stress_Hard-device}
\sigma=\frac{1}{k_{\infty}}
\left[ e_G- \frac{\sum_i k_i w_i}{N}\right]~.
\end{equation}

The stress \eqref{eq.bar_e} can be alternatively expressed as a function of $\bar{e}$ as follows:
\begin{equation}
\label{eq.Stress_bare}
\sigma=\frac{1}{k_\infty}\left[\bar{e}-\frac{\sum_i k_i w_i}{N}\right]~.
\end{equation}
From this expression, it becomes clear that the stress consists of two contributions: a contribution proportional to the deformation of the
sample (first term in the rhs of Eq.~\eqref{eq.Stress_bare}) and a
restoring component (second term in the rhs of
Eq.~\eqref{eq.Stress_bare}) associated with the configuration $\w$ of
snap-springs. Note that Eq.~\eqref{eq.Stress_bare} reduces to Eq.~\eqref{eq.Stress_Hard-device} in the hard-device limit when $\bar{e}=e_G$.

\begin{figure}
\centering
{\includegraphics[width=7.5cm]{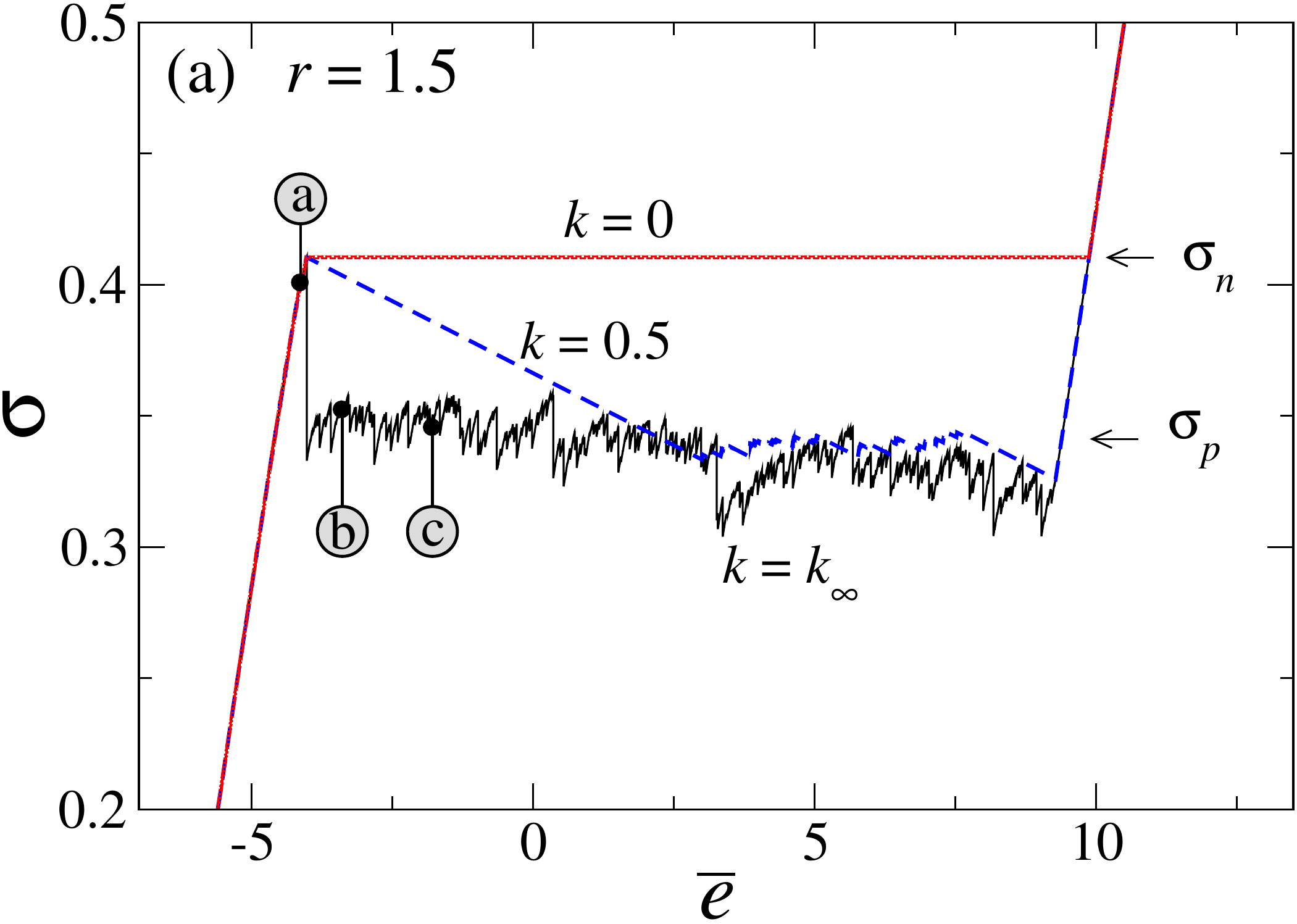}}\\
{\includegraphics[width=7.5cm]{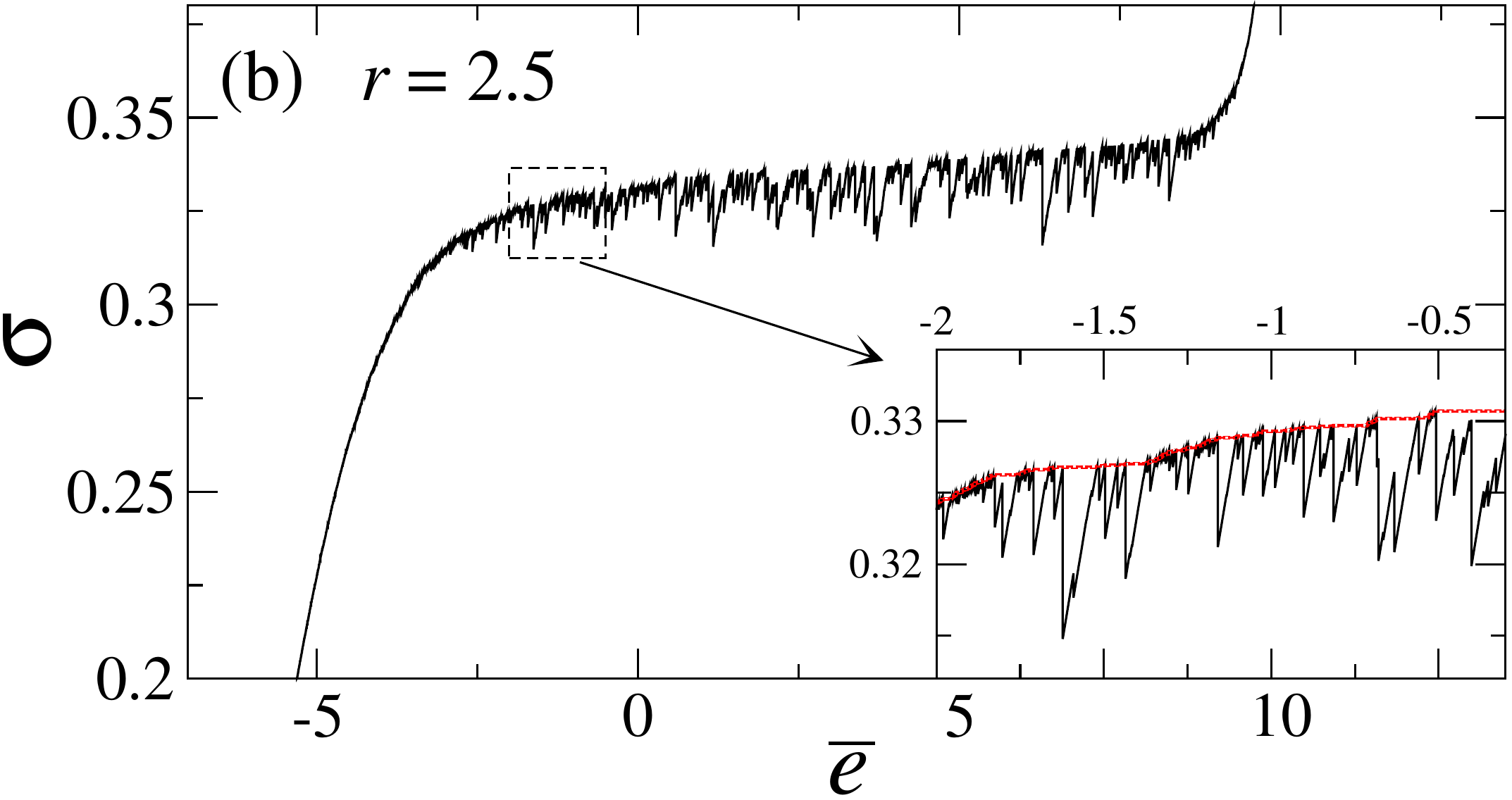}} 
\caption{ \label{Stress-Strain_3D} Stress-strain curves obtained by increasing $e_G$ in systems of size $L=64$ with degree of disorder (a)
  $r=1.5<r_o$ and (b) $r=2.5>r_o$. The
  three curves plotted in (a) correspond to effective stiffness $k=0$ (dotted line),
  $k=0.5$ (dashed line), and $k=k_{\infty}$ (continuous line). The labels \textcircled{a}, \textcircled{b}, and \textcircled{c} along the continuous line indicate the stress-strain values for the snapshots shown in Fig.~\ref{Snapshots}(a), (b), and (c), respectively. The
  curve in (b) corresponds to a hard device loading with $k=k_{\infty}$. The inset in (b) shows
  the stress-strain curves for $k=k_{\infty}$ (continuous line) and
  $k=0$ (dotted line) in the scale indicated by the dashed square in
  the main plot.}
\end{figure}

Fig.~\ref{Stress-Strain_3D} shows the stress-strain curves for two
values of the disorder. For $k=0$, a second-order phase transition occurs for a value of
the disorder $r_o \simeq 2.2$. Such transition is associated with a critical point at $(r_o,\sigma_o)=(2.2,0.34)$ which is equivalent to the OD critical point in the 3D-RFIM~\cite{Sethna_review2004,PerezReche2003,PerezReche2004RFIMField}. In the low disorder regime ($r<r_o$) the system exhibits snap behaviour marked by a macroscopic discontinuity of the strain $\bar{e}$  associated with an infinite avalanche. Such avalanche occurs at a nucleation stress $\sigma_\text{n}$ which is a decreasing function of $r$~\cite{PerezReche_PRL2008}.
In the high disorder regime ($r>r_o$, Fig.~\ref{Stress-Strain_3D}(b)), the transition proceeds through a sequence of small avalanches characteristic of pop behaviour. A disorder-induced transition of this type can indeed be inferred from the results in Ref.~\cite{Wang-Ren_ActaMat2014} for impurity doped martensites. The martensite and strain-glass phases in doped martensites would correspond to the regimes with low and high disorder predicted by the RSSM.

The stiffness of the loading device does not play a significant role for systems
in the pop regime (see the inset in Fig.~\ref{Stress-Strain_3D}(b)). In
contrast, the snap behaviour in the low disorder regime is modified in a non-trivial manner for
$k>0$. As shown
in Fig.~\ref{Stress-Strain_3D}(a), the transformation starts at the stress $\sigma_\text{n}$ as in the case with $k=0$ but then decreases linearly with $\bar{e}$
(cf. Eq.~\eqref{eq.Stress}). More precisely, the behaviour of systems with $k$  smaller than a certain value
$k_\text{p}(r)$, the behaviour is similar to that observed for $k=0$
in the sense that the system is fully transformed in a single snap avalanche to the
 branch with $f_+=1$. In contrast, for $k>k_\text{p}(r)$, the system reaches a
stable branch with $0<f_+<1$ and the transformation then proceeds along
a saw-like path with lower values of the stress, $\sigma_\text{p}$, until the saturation branch with $f_+=1$ is reached. Sect.~\ref{SubSec:Nucleation-propagation} shows that the saw-like path corresponds to a propagation regime
where the phase transformation is dominated by the growth of a single
domain of the new phase.

The nucleation peak predicted by the RSSM at some stress $\sigma_\text{n}$ is indeed observed experimentally in mechanically-driven shape-memory alloys~\cite{Christian2002v,Bonnot2007}. The existence of such peak suggests that the stability limit of equilibrium branches for the RSSM has a re-entrant behaviour in the space $(\bar{e},\sigma)$. This behaviour is similar to that observed in the magnetisation-driven RFIM~\cite{Illa2006a,Illa2006b} and is reminiscent of the re-entrant behaviour of the boundaries of the region of typical\footnote{Typical states have magnetisation $m$ that can be represented by an exponentially large number of microsopic spin configurations, $\bf{s}$.} states reported for the RFIM~\cite{PerezReche_PRB2008,Rosinberg_JSTAT2009}.

Extending the conclusions of a recent study of the spinodal transition in the zero-temperature RFIM~\cite{Nandi-Biroli-Tarjus_Arxiv2015} to the RSSM studied here, one would expect the nucleation peak to be a finite-size effect disappearing  in the thermodynamic limit ($L\rightarrow \infty$). The version of the RSSM studied in this section is however highly simplified and the size-dependence of the nucleation peak in more realistic settings with, e.g., a long-range and anisotropic $\J$, remains to be studied. In principle, an argument based on the RFIM or the simple version of the RSSM studied does not necessarily imply that the nucleation peak will disappear in the thermodynamic limit for martensites.

\subsection{Transformation mechanisms: nucleation and propagation}
\label{SubSec:Nucleation-propagation}

\begin{figure}
\centering
{\includegraphics[width=13cm]{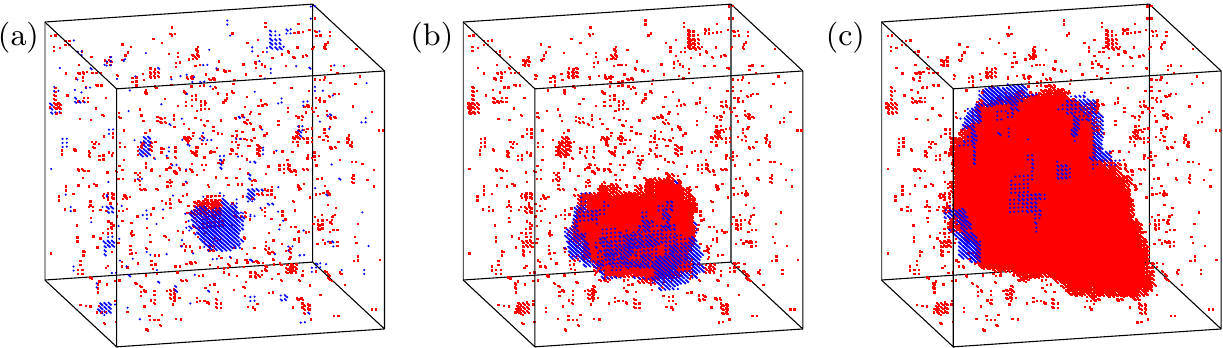}}

 \caption{
   \label{Snapshots} Snapshots showing the transformed domains in the
   system following the stress-strain curve
   of Fig.~\ref{Stress-Strain_3D}(a) for $k=k_{\infty}$. The system
   size is $L=64$ and disorder is $r=1.5$. The red regions correspond to
   snap-springs that have already transformed to the phase $s=+1$ at (a) $f_+=0.005$ ($\bar{e}=-4.14$, \textcircled{a} in Fig.~\ref{Stress-Strain_3D}(a)), (b) $f_+=0.04$
   ($\bar{e}=-3.41$, \textcircled{b} in Fig.~\ref{Stress-Strain_3D}(a)), and (c) $f_+=0.2$ ($\bar{e}=-1.79$, \textcircled{c} in Fig.~\ref{Stress-Strain_3D}(a)).  
Blue regions
   show the transformed snap springs if the driving is slightly
   increased from the value corresponding to the configurations in
   red. Snapshot (a) illustrates the transformation mechanism in the
   nucleation regime before a propagating domain starts growing. Under a
   small increment of the driving, the transformation activity is spatially sparse. 
The propagation regime is illustrated by snapshots
   (b) and (c) where the snap-springs in blue show that \emph{only} the
   propagating domain grows when increasing the driving.}
\end{figure}

To illustrate the effect of the loading stiffness on the
transformation mechanisms of the system, we consider the
particular case with $k=k_{\infty}$.
When increasing $e_G$ as in Fig.~\ref{Stress-Strain_3D}(a), the
new phase nucleates in multiple isolated snap-springs for values of
the stress smaller than $\sigma_\text{n}$ (see the snapshot in
Fig.~\ref{Snapshots}(a)). At $\sigma_\text{n}$, one of the nuclei starts
growing in a process reminiscent of the infinite avalanche occurring
for $k=0$. Such propagating domain is unique because the probability that two or more domains become
unstable at the same stress, $\sigma_\text{n}$, is
zero since $\sigma_\text{n}$ is a real number. 

During the growth of the propagating domain, the stress relaxes at constant driving, $e_G$, until a stable branch is reached. At this point, the new phase stops growing and $e_G$ is increased again. The system evolves elastically (i.e. snap-springs do not change their energy well) along the reached equilibrium branch  until the
stability limit of such branch is reached. At this point, the transformation  resumes. Figs.~\ref{Snapshots}(a) and (b) show that the
transformation proceeds by the intermittent propagation of the
boundary of the propagating domain. The domains of the new
phase other than the propagating domain that nucleated before reaching $\sigma_\text{n}$ were stable for $\sigma=\sigma_n$ and thus remain stable during the
propagation regime which occurs at lower values of $\sigma$. Most of these
domains are absorbed by the propagating domain.

The evolution of the system in the propagation regime consists of a sequence of pinning-depinning (PD) transitions of the boundary of the propagating domain (henceforth referred to as the propagating front). The propagation stress $\sigma_\text{p}$ corresponding to the upper limit of stability of each branch plays the role of a critical force for depinning of the propagating front.

Extrapolating the arguments of Ref.~\cite{Nandi-Biroli-Tarjus_Arxiv2015} to our model, one can argue that droplets of the new phase can in principle nucleate and start growing at the propagation stress, $\sigma_\text{p}$ (i.e. can nucleate and grow before the driving reaches a larger nucleation stress, $\sigma_\text{n}$). The nucleation of such domains is however very unlikely and they are only expected to be frequent for very large systems where there are many possible nucleation events. If one of such nuclei leads to a rare droplet able to grow at the propagation stress, $\sigma_\text{p}$, the system will transform without a nucleation peak at $\sigma_{\text{n}}>\sigma_{\text{p}}$.

\subsection{Universality classes of avalanches}

The results of the previous section show that the transformation mechanisms strongly depend on the stiffness of the loading device, being nucleation-dominated for soft devices ($k<k_{\text{p}}(r)$) and propagation-dominated for harder devices ($k>k_{\text{p}}(r)$). The avalanche behaviour is different depending on the transformation dynamics. For $k<k_{\text{p}}(r)$, the model displays an OD transition between pop and snap regimes which belongs to the universality class of the zero-temperature RFIM with nucleation dynamics. For hard enough loading ($k>k_{\text{p}}(r)$), the system self-organizes to the QEW universality class for driven interfaces; in the RSSM, the interface corresponds to the phase boundary of a propagating domain. Interestingly, the driving-induced crossover between these two transformation mechanisms and critical behaviours was first proposed theoretically in Ref.~\cite{PerezReche_PRL2008} and then observed experimentally~\cite{Vives_AE2009,Vives_PRB2011_ImagingMT}.

\begin{figure}
{\includegraphics[width=6.5cm]{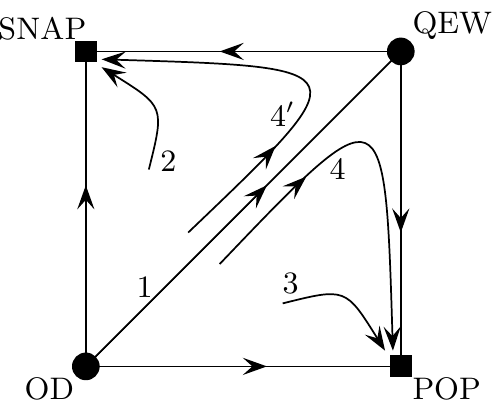}}
 \caption{
   \label{RG_RSSM_Mechanically-Driven_PRL2008} Schematic RG flow for the RSSM model. Separatrix 1 is the QEW universality class manifold which indicates the RG flow from the neighborhood of the OD fixed point to the QEW fixed point. The RG-flow towards SNAP and
  POP regimes is indicated by arrows 2 and 3, respectively. Lines 4 and 4'
  correspond to systems which display QEW critical exponents with supercritical or subcritical cut-offs for $k<k_{\text{p}}(r)$ and $k>k_{\text{p}}(r)$, respectively. [From \cite{PerezReche_PRL2008}, Fig. 4(b), pg. 230601-4].}
\end{figure}

\begin{figure}
{\includegraphics[width=9.0cm]{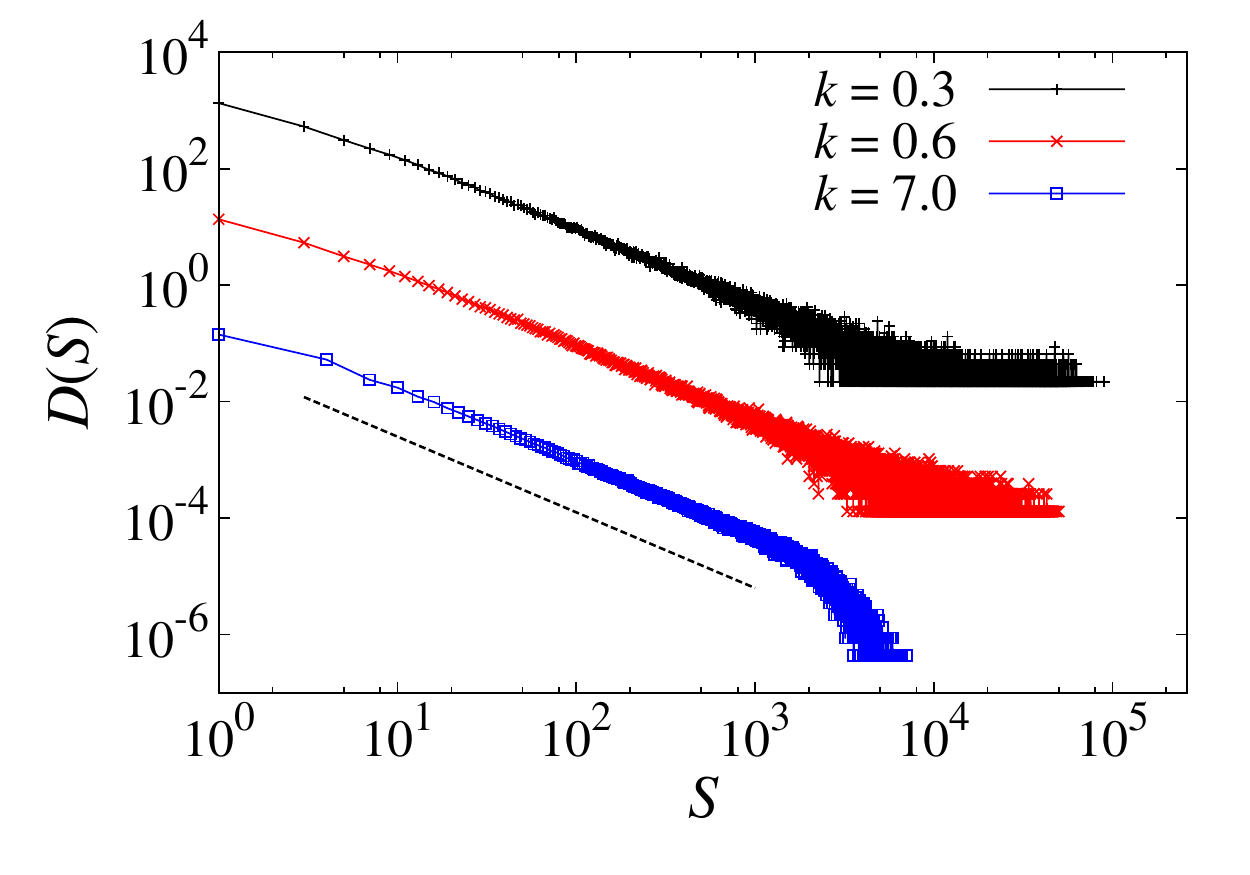}}
 \caption{
   \label{Ds_RSSM_Mechanical_L64} Log-log plot of the distribution of sizes of avalanches exhibited by systems of size $L=64$ and disorder $r=1.5$ in the propagation regime. Different symbols correspond to different values of the stiffness $k$, as marked in the legend. The curves for $k=0.6$ and $k=7$ have been displaced vertically for clarity. The dashed line indicates a power-law $D(S) \sim S^{-\tau_p}$ with $\tau_p=1.3$.}
\end{figure}

The variety of nonequilibrium regimes observed in the mechanically-driven RSSM can be explained using Renormalization Group (RG) arguments~\cite{PerezReche_PRL2008}. The RG studies the way physical systems change under coarse-graining in order to understand the behaviour at large scales~\cite{Wilson1983,Goldenfeld1992,Cardy1996,Sethna2001}. Fixed points in the model parameter space remain invariant under the RG transformation and correspond to systems that remain invariant under coarse-graining. Under the RG transformation, systems flow towards a fixed point which dictates their behaviour at large scales. Four fixed points where assumed for the RSSM, see Fig.~\ref{RG_RSSM_Mechanically-Driven_PRL2008}. Snap an pop behaviours are associated with trivial, fully attractive fixed points where the correlation length between snap-springs vanishes (these points are analogous to those of bulk phases in thermodynamics equilibrium~\cite{Goldenfeld1992}). In contrast, scale-free responses associated with OD and QEW universality classes are dictated by critical fixed points characterised by infinite correlation length between snap-springs\footnote{See \cite{Handford_JSTAT2012} for an explicit calculation of the spin-spin correlation function near the OD transition}.  The OD critical response is associated with a fully repulsive critical point which can be reached only by tuning all four parameters of the model: $\sigma=\sigma_o$, $r=r_o$, $k=0$ and $L^{-1}=0$. In contrast, QEW is a saddle point with a stable manifold which governs the large scale behaviour of the systems with $r < r_o$, $\sigma=\sigma_p(r)$, $k=0$, and $L^{-1}=0$; the corresponding systems lay on the critical manifold connecting the OD and QEW points [arrow 1 in Fig.~\ref{RG_RSSM_Mechanically-Driven_PRL2008}]. Note that $k=0$ is a necessary condition for pure critical behaviour of any type since $k>0$ introduces a restoring force that prevents avalanches from growing indefinitely. This leads to a truncated power-law distribution for the avalanche sizes with a cut-off being increasingly pronounced for increasing $k$ (see  Fig.~\ref{Ds_RSSM_Mechanical_L64}). This implies that finite systems which require $k \geq k_p(r) \geq 0$ to reach a propagation regime are close to the critical QEW manifold [arrow 1 in Fig.~\ref{RG_RSSM_Mechanically-Driven_PRL2008}] but eventually flow towards the POP fixed point under the RG transformation [arrow 4' in Fig.~\ref{RG_RSSM_Mechanically-Driven_PRL2008}]. The situation in the thermodynamic limit can be different if rare droplets are able to grow by front propagation at $\sigma_p$~\cite{Nandi-Biroli-Tarjus_Arxiv2015}. In this case, there is no nucleation peak and the system can reach the propagation regime for arbitrarily small $k$ so that the QEW criticality could be exactly reached.

\section{Conclusions}

In this chapter, it has been highlighted the idea that a complete description of  the deformations occurring during martensitic transformations requires going beyond the usual assumption that the energy of solids is invariant under certain point group elements~\cite{Ericksen_ArchRationalMechAnal1980,Folkins1991,PitteriZanzotto2003,Conti2004,Bhattacharya2003,Bhattacharya2004}. Indeed, crystallographic point groups are just finite subgroups of a global, infinite and discrete symmetry group which includes nonorthogonal and shearing deformations. Accounting for such deformations is crucial to model the emergence of defects such as dislocations induced by many martensitic phase changes. In essence, transformation-induced defects can be viewed as an inherent feature of many martensites which is built in their space of possible deformations. We have presented snap-spring (i.e. pseudo-spin) models which effectively account for such effects and provide an explanation for the emergence of scale-free avalanche behaviour after a training period. Such models account for both elastic and non-elastic properties of solids and this makes them ideal to study the interplay between defects and the phase transition in both thermally and mechanically driven martensitic transformations. 

Using a simple version of a random snap-spring model, it has been shown that criticality in mechanically-driven transformations depends on the stiffness of the loading device. The analysis presented here has only considered weak transformation and assumed short-range positive-definite interactions. Such simplification allowed us to compare with well-studied random-field models but a complete description of mechanically-induced transformations will require using long-range interactions and allowing for transformation-induced defects.

The role of transformation-induced defects was analysed for thermally-driven transformations. In this case, the evolution to criticality was initially interpreted in terms of the self-organized criticality (SOC) paradigm~\cite{PerezReche2007PRL,PerezReche_CMT2009}. This interpretation was however challenged by experiments on well-trained martensites which suggested that criticality requires tuning the driving parameter (temperature) to a critical value~\cite{Chandni_PRL2009}. These results would support the existence of a critical point of the type displayed by random-field models rather than self-organized criticality. In a recent work~\cite{Perez-Reche_PRB2016} we extended the random-snap spring model reviewed here to propose an explanation that unifies the two seemingly conflicting interpretations proposed in Refs.~\cite{Ahluwalia2001,PerezReche2007PRL,PerezReche_CMT2009,Chandni_PRL2009}.  More explicitly, we identified a critical manifold in the temperature-disorder space of martensites where they are marginally stable~\cite{Muller-Wyart_annurev-conmatphys2015}. The evolution of disorder during the phase transition allows the system to approach the critical manifold at a critical temperature without extrinsic tuning of disorder.  This mechanism is reminiscent of the criticality paradigm proposed in Ref.~\cite{Gil-Sornette_PRL1996} which postulated that a suitable coupling between driving and order parameters can lead to robust criticality. Our model predicts that a coupling of this type between the temperature and slip disorder explains the robustness of criticality in martensitic transformations.

Following the tradition of statistical mechanics, the models reviewed in this chapter are intended to capture generic properties of martensitic transformations. They are based on a number of simplifying hypotheses which include neglecting thermally activated effects or assuming infinitely slow driving fields. Thermal fluctuations are indeed a secondary factor for many shape-memory alloys~\cite{PerezReche2001}. The very fact that avalanches are observed as separated events in slowly driven systems indicates that thermally activated events are not frequent. In spite of that, some thermally activated events can occur~\cite{PerezReche2001} and spin models can be extended to study their effect on avalanches. This was done in Ref.~\cite{PerezRechePRL2005} which focused on systems with weak thermal fluctuations and also investigated the effects of finite driving rates on avalanches. Both finite driving and thermal fluctuations promote the merging of avalanches that would be detected as separated events in athermal, quasistatically driven systems. As a consequence, the exponent for the distribution of avalanche sizes is smaller when thermal fluctuations are active and/or the driving is not quasistatic. In systems with stronger thermal fluctuations, individual avalanches are strictly speaking undetectable but bursts of transformation activity can still be distinguished. Molecular dynamics simulations show that the energy of such bursts deviates from the power-law behaviour as thermally activated effects become stronger~\cite{Salje_PRB2011}. The mechanisms responsible for such deviations are still not fully understood.

Understanding the consequences of relatively strong thermal fluctuations within the framework of spin and/or snap-spring models is an interesting  challenge for future studies. Another interesting task would consist in extending the proposed random-snap spring model to incorporate more realistic interactions and study generic phase transition paths in 2D and 3D systems. Such extensions will lead to a better understanding of the factors responsible for training effects and universality classes of avalanche dynamics in realistically complex materials. The random-snap spring model sets a good basis to achieve these goals.

\vskip10pt
{\bf Acknowledgements}
The author is grateful to Lev Truskinovsky and Giovanni Zanzotto for an enlightening  collaboration on the topics covered in this chapter. The author is also grateful for insightful discussions and/or collaboration with a number of researchers including Eduard Vives, Antoni Planes, Llu{\'\i}s Ma{\~n}osa, Jordi Ort{\'\i}n, Carles Triguero, Eckhard Salje, Sergei Taraskin, Stefano Zapperi, Turab Lookman and Avadh Saxena.



%

\end{document}